\documentclass{emulateapj}                 

\usepackage{graphicx}
\usepackage{epsfig}
\usepackage{amsmath}
\usepackage{subfigure} 
\usepackage{hyperref}
\usepackage{amssymb}
\usepackage{lscape} 
\usepackage{apjfonts}
\usepackage{latexsym}
\usepackage{parskip}

\newcommand{\Msun}{\text{M}_\odot}
\newcommand{\Zsun}{\text{Z}_\odot}
\newcommand{\cc}{cm $^{-3}$}

\shorttitle{Fragmentation in dusty low-metallicity star forming halos}
\shortauthors{G. Meece et. all}

\begin{document}

\title{Fragmentation in dusty low-metallicity star forming halos}

\author{Gregory R. Meece}
\affil{Department of Physics and Astronomy, Michigan State University,
East Lansing, MI 48824, USA}
\author{Britton D. Smith}
\affil{Department of Physics and Astronomy, Michigan State University,
East Lansing, MI 48824, USA}
\author{Brian W. O'Shea}
\affil{Lyman Briggs College, Michigan State University, East Lansing, MI 48825, USA}
\affil{Department of Physics and Astronomy, Michigan State University,
East Lansing, MI 48824, USA}

\begin{abstract}
The first stars in the universe, termed Population III, are thought to have been very massive compared to the stars that form in the present epoch. As feedback from the first generation of stars altered the contents of the interstellar medium, the universe switched to a low-mass modeof star formation, which continues in the high metallicity stars formed in the present era. Several studies have investigated the transition between metal-free and metal-enriched star formation, with tentative evidence being found for a metallicity threshold near $10^{-3.5}~\Zsun$ due to atomic and molecular transitions and another threshold near $10^{-5.5}~\Zsun$ due to dust. In this work, we simulate the formation of stars in idealized low-metallicity halos using the AMR code Enzo. We conduct several simulations of $10^6~\Msun$ and $10^7~\Msun$ halos in which the metal content, initial rotation, and degree of turbulence are varied in order to study the effect of these properties on gas fragmentation over a range of densities. We find tentative support for the idea of a critical metallicity, but the effect of varying metallicity on the gas we observe is not as dramatic as what has been reported in earlier studies. We find no clear relation between the initial spin or the initial level of turbulence in the halo and the final properties of the gas contained therein. Additionally, we find that the degree to which the Jeans length is refined, the initial density profile of the gas, and the inclusion of deuterium chemistry each have a significant effect on the evolution and fragmentation of the gas in the halo -- in particular, we find that at least 64 grid cells are needed to cover the Jeans length in order to properly resolve the fragmentation.
\end{abstract}

\section{Introduction}\label{section:introduction}
It is well established that the very early universe contained only trace amounts lithium and essentially no other elements heavier than hydrogen and helium \citep{steigman_2007, wagoner_1973}. After Big Bang nucleosynthesis, virtually all heavy elements are synthesized in stars. It follows that the first stars, termed Population III stars, must have been free of heavy elements. However, observations have yet to identify any of these metal free stars \citep{ryan_1996, beers_2005, caffau_2012, yong_2013}. Additionally, observations of Lyman-$\alpha$ systems reveal that even low density gas at high redshifts is contaminated by heavy elements, indicating significant enrichment by earlier generations of stars \citep{cowie_1998}. This leads to the conclusion that the first stars were massive and short-lived \citep{barkana_2001, ripamonti_2004, bromm_2004, glover_2005, norman_2010}.\par

To produce a stellar initial mass function (IMF) consisting of mostly high mass stars, the process of star formation in the primordial universe must have differed substantially from modern day star formation. Stars form when over-dense clouds of gas radiate energy and collapse due to self gravity. Density perturbations larger than the Jeans length will tend to collapse faster than the surrounding gas. If the gas is able to efficiently radiate energy as it collapses, such that the Jeans Mass decreases with increasing temperature, the gas will continuously fragment. Thus, the final mass of the protostellar cloud will be set by the Jeans mass at the point where the gas can no longer cool efficiently. The initial stellar mass will be set by the size of the protostellar cloud and accretion, although the details of this process are quite complicated, and the Population III IMF is highly uncertain as a result \citep[e.g.]{tan_2004, mckee_2008, norman_2010, clark_2011, greif_2011}.\par

The ability of the gas cloud to cool will be set by the micro-physics of the gas. In the local universe, rotational lines in CO and line cooling from  CI and OI are primarily responsible for cooling \citep{omukai_2000}, and are able to lower the temperatures of star forming clouds to around 10 K. In the early universe, however, the only significant sources of cooling were H$_2$ and HD molecules. Due to the lack of a permanent dipole in H$_2$, the rotational energy levels are relatively widely spaced, and rotational transitions are not able to cool the gas below a temperature of around 200K \citep{galli_1998}. While HD is a more effective coolant owing to a permanent dipole moment, the low initial fraction of deuterium prevents a high HD fraction from forming, typically preventing HD from contributing to the total cooling as much as H$_2$ \citep{galli_2002}. If heavy elements are present, the gas will be able to cool faster and to lower temperatures than is possible in primordial gas. Metals in the form of dust will be able to cool the gas through thermal radiation \citep{omukai_2005, schneider_2010}. Dust can also serve as a catalyst for H$_2$ formation, providing an additional source of cooling.\par

Many authors have studied the transition from metal-free to metal-enriched star formation using idealized models. While the initial conditions of these models are necessarily less accurate than those of cosmological simulations, their fully specified nature allows one parameter to be varied at a time, facilitating our ability to isolate and understand the effects of individual physical processes. Over the past decade, idealized simulations have explored more of the relevant physical and chemical processes underlying star formation in the early universe, resulting in tunable models that more accurately capture the conditions of primordial and low metallicity star formation. \citet{bromm_2001} modeled a $2.0 \times 10^6~\Msun$ top-hat overdensity collapsing at $z=30$ and found the first evidence of a `critical metallicity' of approximately $5 \times 10^{-4}~\Zsun$. \citet{omukai_2005} has studied the thermodynamics of collapsing primordial and low metallicity gas using one-zone models. More recently, a series of works \citep{jappsen_2007a, jappsen_2007b, jappsen_2009a, jappsen_2009b} modelled the collapse of a hot, ionized gas which had been allowed to relax to hydrostatic equilibrium within an NFW potential \citep{navarro_1996} before cooling was turned on. This group concluded that there is no clear critical metallicity, and that fragmentation is more dependent on the choice of initial conditions.\par

Several works \citep{omukai_2000, omukai_2005, schneider_2006, schneider_2010} have focused on the effects of dust cooling on fragmentation in low metallicity clouds. Dust cooling is typically effective at densities above $n_H=10^{10}~$\cc, where the gas and dust temperatures are coupled. These studies have found evidence of a lower metallicity threshold around $10^{-5.5}~\Zsun$ due to dust cooling when dust is included in simulations.\par

In this work, we extend the study of the transition from metal-free to metal-enriched star formation by using an idealized model based on the results of cosmological simulations. Our model uses several parameters to set the metallicity, chemistry, and the shape of the initial density, temperature, and rotational profiles, as well as allowing for different levels of turbulence and different halo masses. In Section \ref{section:method}, we discuss our simulation code and the initial setup of our star-forming halos in detail. In Section \ref{section:evolution_of_fiducial_model}, we provide an overview of the evolution of our fiducial model. Section \ref{section:varying_refinement_criteria} discusses the effects of varying our refinement criteria and establishes the criteria necessary to adequately resolve the collapse. In Section \ref{section:exploration_of_parameter_space}, we discuss the evolution of our model for different points in the parameter space of metallicity, rotation, turbulence, and dust. In Section \ref{section:discussion}, we discuss the assumptions in our simulations that may influence our results, including the effects of deuterium chemistry, the shape of the initial density profile, and the validity of our chemical model. We summarize and conclude in Section \ref{section:conclusion}.\par

\section{Method} \label{section:method}

\subsection{The Simulation Code and Included Physics} \label{section:enzo}
We model the collapse of the halo using the Eulerian adaptive mesh refinement code Enzo \citep{oshea_2004, norman_2007, enzo_2013}. The hydrodynamics are calculated using the piecewise parabolic method of \citet{colella_1984}. In order to ensure conservation of mass within our simulation, we employ periodic boundary conditions for the gas. To calculate the gravitational potential, we assume isolated boundary conditions. In addition to the self gravity of the baryons, we calculate the gravitational potential of a static NFW halo. Each simulation is initialized with a top level grid resolution of $128^3$ cells and is refined during setup. Although we do not use comoving coordinates in this work, we assume a redshift of $z=20$ where necessary during initialization, and all distances in this paper in are in proper parsecs at that redshift. Each halo is placed in the center of a box with a proper size of 2000 pc per side. During initialization, we require that the inner 100 pc be covered by four levels of grid refinement, giving a maximum spatial resolution of 0.977 pc at the beginning of the simulation. As the virial radius of the dark matter halo (taken to be the edge of the sphere) is an order of magnitude smaller than the box size, the effects of boundary conditions on the evolution of the halo should be negligible.\par

\subsubsection{Refinement Conditions}\label{section:refinement_criteria}
We employ four criteria for determining when to refine grid cells. In all simulations, refinement is carried out by subdividing a grid cell by a factor of 2 along each dimension, thus into 8 equal-sized cells. Lagrangian refinement would therefore require that we refine a grid cell whenever the enclosed mass exceeds the average mass in one top level grid cell by a factor of eight. To better understand the evolution of the densest regions, we impose super-Lagrangian refinement by refining whenever cell mass exceeds

\begin{equation} \label{equation:mass_refinement}
   M_{cell} > M_{top} \times 8^{-0.3 \cdot l}
\end{equation}

Where $l$ is the current refinement level.\par

Our second refinement condition splits a cell whenever the local cooling time, $t_{cool}$, is shorter than the sound crossing time of the cell, $\Delta x/c_s$. This requirement is necessary to justify our assumption that the gas is thermodynamically stable at scales smaller than the grid resolution. Thirdly, we refine when the size of a cell is larger than some fraction of the local Jeans length, when $\lambda_J < N_J \Delta x$, where $\lambda_J$ is the Jeans length (calculated in that cell) and $N_J$ is the number of cells over which the Jeans length must be refined. Unless otherwise noted, we require that the Jeans length be resolved by at least 64 cells at all times, e.g. $N_J=64$. In Section \ref{section:varying_refinement_criteria}, we discuss a series of tests to determine the minimum value of $N_J$ necessary in order to accurately model the fragmentation of the collapsing halo. Finally, we require that the region within a cube with side length 100 pc centered on the center of the sphere is always covered by a spatial resolution of less than 1 pc. This criterion ensures that the conditions in the inner region of the halo, as defined by our initial setup, are accurately captured by the mesh. We allow the simulation to dynamically refine using up to 25 levels of grids, for a maximum spatial resolution of 0.0962 AU.\par

\subsubsection{Chemistry Model} \label{section:chemistry_model}
Our chemistry model follows the non-equilibrium reactions for 12 primordial chemical species (H, H$^+$, He, He$^+$, He$^{++}$, e$^-$, H$_2$, H$_2^+$, H$^-$, D, D$^+$, and HD) \citep{anninos_1997, abel_1997} and includes H$_2$ chemistry with three body $H_2$ formation \citep{abel_2002} and H$_2$ formation heating \citep{turk_2009}. In addition, we use the cooling model of \citet{smith_2008} to track cooling from metals in simulations where metals are present. Unlike the chemical model used for primordial species, the metal cooling model does not explicitly track the abundance of individual metal ions. Instead, we use data generated by the photoionization code CLOUDY (see \citet{ferland_1998}) to calculate metal cooling rates for a wide range of densities. Throughout this paper, we assume a scaled solar abundance pattern. The validity of this choice is discussed in Section \ref{section:limitations}. In Section \ref{section:choice_of_chemistry_model}, we discuss the effects of using a reduced chemical model that does not include deuterium chemistry.\par

\subsubsection{Dust Model} \label{section:dust_model}
The presence of dust grains alters the thermal state of the gas by providing a very efficient channel for H$_{2}$ formation and through heat transfer via elastic collisions with the gas. Dust grains cool via continuum thermal emission and are heated by incident radiation. Currently, we only consider incident radiation from the CMB, but in principle an additional heating term can be easily added. The rates of H$_{2}$ formation on grain surfaces and heat exchange with the gas are dependent on the grain temperature, $T_{gr}$, which we assume to be in instantaneous equilibrium. The implementation employed here closely follows that of \citet{omukai_2000} and \citet{omukai_2005}. We calculate the grain temperature by solving the heat balance equation given by

\begin{equation} \label{equation:dust_temperature}
   4 \sigma T_{gr}^{4} \kappa_{gr} = \Lambda_{gas/grain} + 4 \sigma T_{rad}^{4} \kappa_{gr},
\end{equation}

where $\sigma$ is the Stefan-Boltzmann constant and $T_{rad}$ is the radiation temperature, specifically the CMB temperature here.  The rate of heat exchange between the gas and dust per unit dust mass, $\Lambda_{gas/grain}$, is given by 

\begin{multline} \label{equation:dust_gas_heat_exchange}
   \Lambda_{gas/grain} =\\
   1.2\times10^{-31}~\frac{n_{H}^{2}}{\rho_{gr}} \left(\frac{T}{1000 K}\right)^{1/2} (1 - 0.8 e^{-75 / T}) (T - T_{gr})\\
   ~\textrm{erg~s$^{-1}$~g$^{-1}$}
\end{multline}

\citep{hollenbach_1989}, where n$_{H}$ is the H number density and $\rho_{gr}$ is the dust mass density.  We assume that as metallicity increases, dust remains a constant fraction of the metallicity.  We adopt the piecewise polynomial approximation of the grain opacity of \citet{dopke_2011}, given by

\begin{equation}
   \kappa(T_{gr}) \propto \left\{ \begin{array}{ll}
      T_{gr}^{2} & \textrm{, $T_{gr}$ $<$ 200~K,}\\
      constant & \textrm{, 200~K $<$ $T_{gr}$ $<$ 1500~K,}\\
      T_{gr}^{-12} & \textrm{, $T_{gr}$ $>$ 1500~K},
   \end{array} \right.
\end{equation}

with a normalization of $\kappa_{gr}(T_{gr} = 200~K) = 16$ cm$^{2}$ g$^{-1}$ \citep{pollack_1994, omukai_2000}.  The steep power law index for $T > 1500$ K mimics the effect of grains melting.  We take the exact form of the rate for H$_{2}$ formation on grains given in \citet{omukai_2000}, which is derived from the work of \citet{tielens_1985}.  We include the heating/cooling from H$_{2}$ formation/destruction following \citet{omukai_2000} and \citet{hollenbach_1979}.\par

\subsection{Initial Conditions}\label{section:initial_conditions}
We model the star forming regions as a spherically symmetric baryonic halo with a turbulent velocity field within a static NFW potential. Our models are empirically motivated by the results of cosmological simulations \citep{oshea_2007, smith_2009} and informed by the one-zone models of \citet{omukai_2005}. Although our simulations are non-cosmological, we assume that the calculation proceeds at a fixed redshift of $z=20$ for the purposes of calculating heating and cooling rates due to the cosmic microwave background. We assume an $\Lambda CDM$ universe with $\Omega_\Lambda = 0.7$, $\Omega_M = 0.3$, and $H_0=70$ km s$^{-1}$ Mpc$^{-1}$ where relevant during the initialization. These parameters are used when calculating the virial radius of the halo, and small variations in cosmological parameters would not have a large effect on our simulations. We assume that the halo is decoupled from the Hubble flow, and do not take cosmological expansion into account during the simulation (which is reasonable, as the halo is overdense enough to be decoupled from the expansion of the universe.) Thus, all distances quoted in this work are in physical (i.e., proper) units.\par

\subsubsection{Dark Matter Halo}\label{section:dark_matter_halo}
The dark matter component of the halo is assumed to reside in an NFW halo \citep{navarro_1996},

\begin{equation}\label{equation:nfw_profile}
   \rho_{DM}(r) = \frac{\rho_c}{\left(r/R_S\right) \left(1+r/R_S\right)^2}
\end{equation}

where $\rho_c$ is equal to four times the density at the virial radius and $R_S$ is the scale radius.\par

The concentration parameter of the halo, defined as

\begin{equation}\label{equation:nfw_concentration_parameter}
   c = \frac{R_{178}}{R_S}
\end{equation}

is set to $c=2$ for our simulations. Here, $R_{178}$ is the virial radius, calculated as the radius at which the average enclosed density is 178 times the critical density of the universe (see \citet{bryan_1998} for more discussion). Our value of $c$ is within the expected range for the halos we are studying, as predicted by \citet{davis_2010}. In this work, the mass of the halo is taken to mean the mass of dark matter within the viral radius of the halo. Due to the low initial baryon density, the total mass of the halo is not substantially higher. We study models with dark matter masses of $10^6~\Msun$ and $10^7~\Msun$, with corresponding virial radii of 153 and 329 pc. These halos are hereafter referred to as the ``low mass'' and ``high mass'' halos, respectively.\par

\subsubsection{Baryon Density and Temperature}\label{section:density_and_temperature_profiles}
\begin{figure*}
   \includegraphics{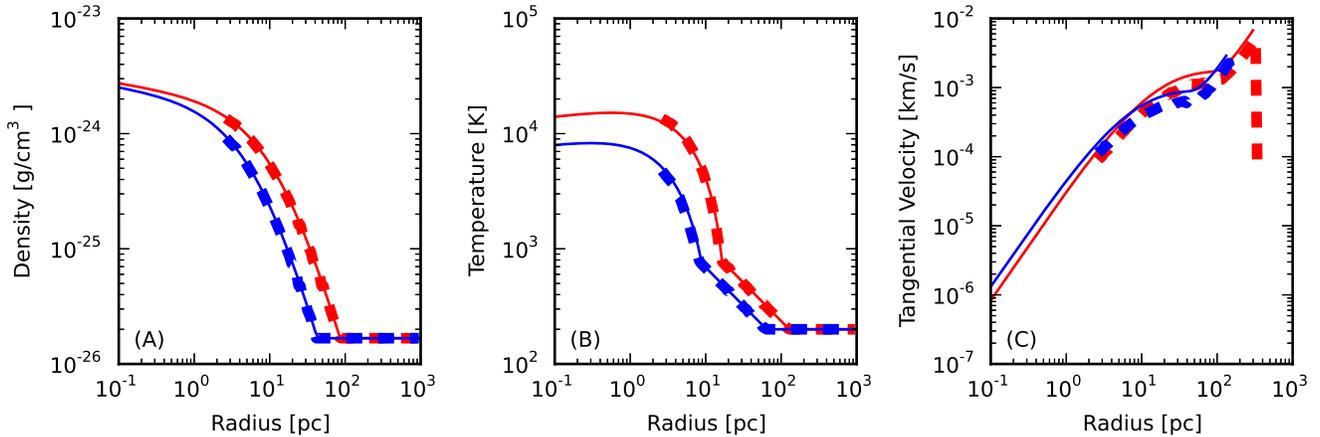}
   \centering
   \caption{The initial conditions of our model are shown for our high and mass fiducial models. Solid lines represent the theoretical values, while the dashed lines are the values realized in our simulation. Panel A shows density as a function of radius. The initial temperature profile is derived by assuming hydrostatic equilibrium in the core and a power law fall off in the envelope, and is shown in Panel B. The rotational velocity, shown in Panel C, is derived by assuming that average angular momentum follows a power law relationship as a function of mass enclosed.}
   \label{fig:initial_conditions}
\end{figure*}

The baryonic component of the halo is modeled by a core in roughly hydrostatic equilibrium with a diffuse envelope. The envelope drops off rapidly until it reaches the background density of $n_H=10^{-2}~$ \cc. The density profile is described by

\begin{equation}\label{equation:density_profile}
   \rho_B(r) = \frac{\rho_B}{\left(r/R_{core}\right)^{\alpha}\left(1+r/R_{core}\right)^{\beta-\alpha}}
\end{equation}

and is shown in the Panel A of Figure \ref{fig:initial_conditions}. For our simulations, we use $\alpha=0.1$ and $\beta=2.5$. These values were chosen by fitting the results of cosmological simulations.\par

For the initial density profile, we use a central baryon number density of $n_H=1~$\cc for both the high and low mass fiducial cases. For the low mass halo, we choose a core radius of $R_{core}=8$ pc and for the high mass case $R_{core}=16$ pc. Our choice of a low initial central density (compared to the dark matter density) is motivated by the desire that the simulation have time to `forget' the details of the initial conditions and reach a stable configuration before collapse sets in. In Section \ref{section:varying_density_profile}, we discuss the results of starting a simulation with a higher initial central density.\par

The temperature profile is calculated by assuming that the gas is in hydrostatic equilibrium within the core and is being adiabatically heated in the envelope. The initial temperature profiles is shown in Panel B of Figure \ref{fig:initial_conditions}.

\subsubsection{Chemistry and Metallicity} \label{section:initial_chemistry}
We initialize the gas in our model to a composition consistent with conditions in the $z=20$ universe for gas that has not been affected by recent star formation. At initialization, all simulations have a uniform electron fraction of $\chi_e=1.69 \times 10^{-4}$, based on calculations performed with the code RECFAST \citep{seager_1999, seager_2000}, and a corresponding HI fraction of $f_{\text{HI}}=0.999831$. The H$^-$ fraction is $f_{\text{H}^-}=10^{-10}$. The initial molecular hydrogen fraction is $f_{\text{H}_2\text{I}}=10^{-4}$. Initial values for D and HD are scaled to the H and H$_2$ values using a D/H mass ratio of $6.8 \times 10^{-5}$.\par

In models where metals are present, we assume a scaled solar abundance of heavy elements. Metallicity is kept uniform throughout the simulation. In simulations where dust is present, it is assumed that the mass fraction of heavy elements in dust is $9.23 \times 10^{-3}$. The effects of our choice of initial chemistry is discussed in Section \ref{section:discussion}. \par

\subsubsection{Velocity Profile}\label{section:initial_velocity_profile}
The halo is given an initial angular momentum distribution characterized by the dimensionless baryonic spin parameter, defined in \citet{peebles_1971} as 

\begin{equation}\label{equation:spin_parameter}
   \lambda = \frac{J |E|^{1/2}}{G M^{5/2}}
\end{equation}

where $J$ is the total angular momentum of the baryons, $E$ is the binding energy of the baryons, $G$ is the gravitational constant, and $M$ is the mass of the baryons. Based on the results of \citet{oshea_2007}, the angular momentum is distributed so that the specific angular momentum as a function of mass enclosed is given by

\begin{equation}\label{equation:angular_momentum_profile}
   |l|(M < r) \propto \left(\frac{M(<r)}{M_{Total}}\right)^{0.9}
\end{equation}

which is scaled by the spin parameter.\par

Gas outside of the virial radius has no initial velocity, and none of the gas has an initial radial velocity before turbulence is added. The dark matter component of the halo is treated as a static potential, and thus has no velocity. The initial rotation profile for our fiducial models is shown in Panel C of Figure \ref{fig:initial_conditions}.\par

\subsubsection{Turbulence}\label{section:turbulence}
The rotational velocity is modified by adding a turbulent velocity field with a power spectrum $P(k) \propto k^{-4}$, suitable for compressible gas \citep{clark_2011}. The turbulent field is generated using the method described in \citet{rogallo_1981}. The turbulent field is applied only within the virial radius of the halo, and is normalized such that the RMS velocity is a specified fraction of the sound speed of the halo, as defined in \citet{barkana_2001}. For consistency, the same turbulence field was used for all simulations.\par

\subsection{Varying the Initial Conditions}\label{section:varying_initial_conditions}
In order to study the importance of different model parameters to the evolution and fragmentation of the gas, we conduct several runs wherein one parameter is systematically varied. For each parameter, we conduct simulations for both the high and low mass halos, as described above. The full list of simulations performed in this work is given in Tables 1-5.\par

\subsection{Clump Finding} \label{section:clump_finding}
To quantify the degree of fragmentation, we run a clump finder on the
central 20 pc of the final output from each simulation.  The clump
finding algorithm, described in detail in \citet{smith_2009} and
implemented in the \texttt{yt}\footnote{http://yt-project.org/} 
simulation analysis toolkit \citep{turk_2011}, works by
identifying topologically disconnected structures in density space.
Following \citet{truelove_1998}, we define a clump or fragment
as ``the mass contained between a local density maximum and the lowest
isodensity surface surrounding only that maximum.''  
\citet{smith_2009} only consider clumps that are strictly 
gravitationally bound, but here we use a modified criterion to include
clumps that are marginally unbound but rapidly cooling, since these
objects will likely become bound in the future.  Clumps are considered
valid if they satisfy the following requirement:
\begin{equation}
KE + TE - \sum_{i} (\Lambda_{i}\ t_{dyn, i}) < PE,
\end{equation}
where KE, TE, and PE are the total kinetic, thermal, and potential
energy of the clump and $\Lambda_{i}$ and $t_{dyn, i}$ are the cooling rate and
dynamical time for reach grid cell that is a member of the clump.

\section{Evolution of the Fiducial Model} \label{section:evolution_of_fiducial_model}
For our fiducial model, we choose high and low mass halos with a metallicity of $Z=10^{-3}~\Zsun$, a spin parameter of $\lambda=0.05$, turbulence normalized to 0.4 times the halo sound speed, and with dust present. This model is chosen to simulate a typical star forming halo at $z=20$, which has not hosted recent star formation (e.g., \citet{oshea_2007, smith_2009}). We choose a metallicity which is in the middle of our range of values, and is near the theoretical ``critical metallicity.'' We mandate that the Jeans length be covered by at least 64 cells at all times by setting $N_J=64$.\par

\begin{figure}
   \includegraphics[width=0.5\textwidth]{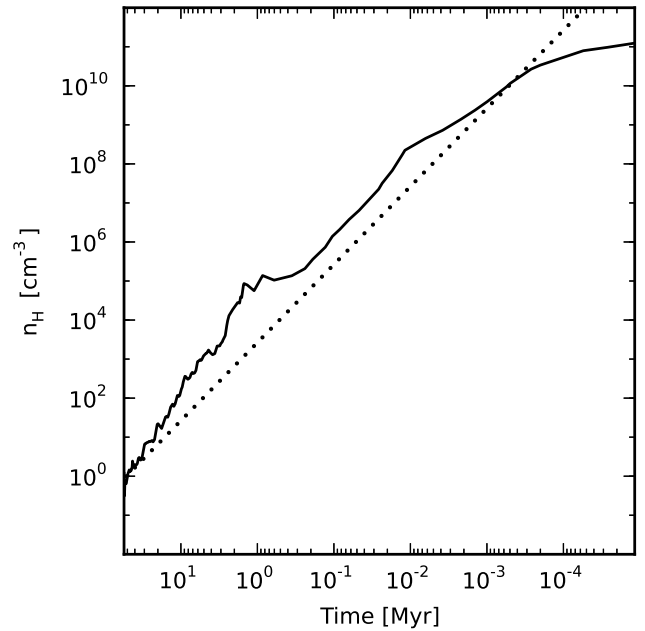}
   \centering
   \caption{The central density of the gas is shown as a function of time. The x axis shows time in millions of years before the last data output. The time from the beginning of the simulation to the last data output is 55.36 million years. The dotted line shows the free fall time of a sphere of gas with a given density. On the left hand side of the plot, the central density is increasing faster than the free fall time scale (as indicated by the slopes of the lines), indicating that the dark matter rather than self gravity is controlling the dynamics of the gas. During the last million years of the simulation, the self gravity of the gas dominates in the center. The gas evolves in free fall until roughly 10,000 years before the end of the simulation, in which pressure support delays further collapse.}
   \label{fig:high_mass_density_evolution}
\end{figure}

We run our simulations until the central density has reached a central hydrogen number density of at least $n_H=10^{10}~$ \cc, which is near the limits of the validity of our cooling model and is approaching the point where our assumption that the gas is optically thin begins to fail. The high mass fiducial halo collapses 55.36 million years after the beginning of our simulation. The evolution of the central density as a function of time is shown in Figure \ref{fig:high_mass_density_evolution}. The collapse begins slowly and accelerates as density increases.\par

\begin{figure*}
   \includegraphics[width=1.0\textwidth]{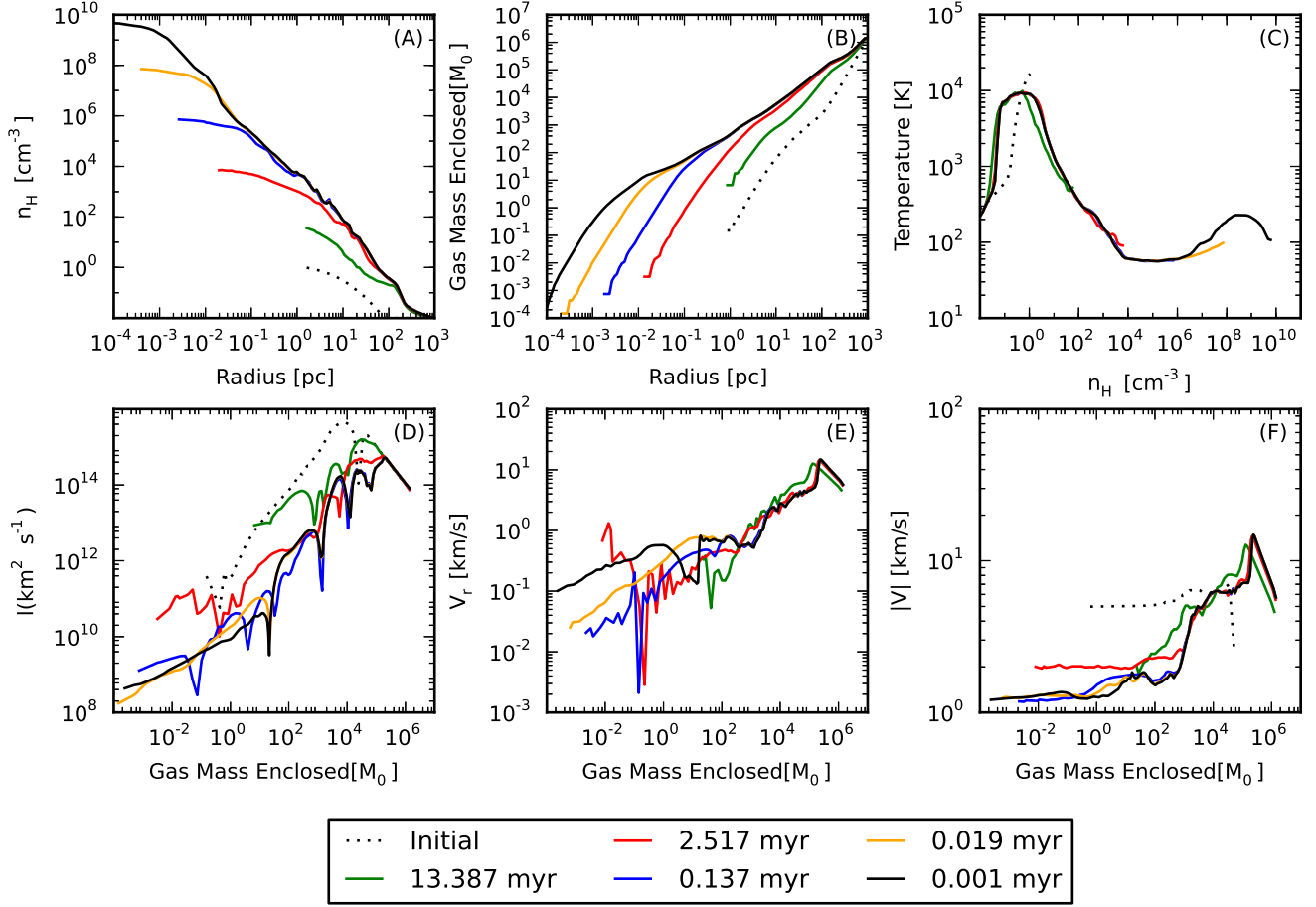}
   \centering
   \caption{The physical state of the gas in our high mass fiducial model is shown for a series of outputs. The green, yellow, blue, red, and black lines show the first outputs in which the central density reaches $10^2$, $10^4$, $10^6$, $10^8$, and $10^{10}$ \cc, respectively. For each output, the legend shows the time remaining until the end of the simulation. Panel A shows spherically averaged gas density as a function of radius, centered on the densest point in the simulation. Panel B shows the total gas mass enclosed as a function of radius. Panel C shows the mass weighted spherically averaged temperature of the gas as a function of density. Panel D shows the mass averaged angular momentum as a function of enclosed mass. For a spherically symmetric collapse with not angular momentum transport, the angular momentum profile would not change with time. The fact that it does indicates that angular momentum is being transported out of the core by turbulence. Panels E and F show the mass weighted spherically averaged radial velocity and velocity magnitude of the gas as a function of mass enclosed. In each panel except Panel E, the initial conditions are represented by a dotted black line. The gas has no initial net radial velocity.}
   \label{fig:high_mass_physical_evolution}
\end{figure*}

When the simulation is initialized, there is a period lasting around 10 million years during which the velocity profile evolves into a steady state. During this time, the details of the initial conditions are wiped out. At this point, the gas in the envelope is collapsing in free fall and is being heated through adiabatic compression, while the gas in the core is pressure supported. An accretion shock forms at the edge of the sphere, near the virial radius. As the gas is decelerated, it is heated to the virial temperature of the halo. As shown in Panel A of Figure \ref{fig:high_mass_physical_evolution}, the collapse evolves self similarly, with the size of the core shrinking as the gas collapses to higher central densities. The velocity profile of the gas as a function of enclosed mass, shown in Panels E and F of Figure \ref{fig:high_mass_physical_evolution}, remains roughly constant, with the gas in the envelope in free fall and the gas in the core collapsing slowly. The gas in the core evolves quasi-statically until near the end of the simulation, at which point the gas is able to efficiently cool and collapses on a free fall timescale.\par

Panel D of Figure \ref{fig:high_mass_physical_evolution} shows the mass averaged angular momentum of the gas as a function of enclosed gas mass, defined as

\begin{equation} \label{equation:specific_angular_momentum}
   l(M)=\frac{\Delta J(M)}{\Delta M}
\end{equation}

where $\Delta M$ is the mass of gas enclosed within a spherical shell and  $\Delta J(M)$ is the total angular momentum of the gas within the shell. With no angular momentum transfer and no external torque, $l(M)$ would stay constant throughout the collapse. In our simulations, $l(M)$ decreases, indicating that angular momentum is being transported outward (relative to the Lagrangian mass coordinate) in the central regions.\par

\begin{figure*}
   \includegraphics[width=1.0\textwidth]{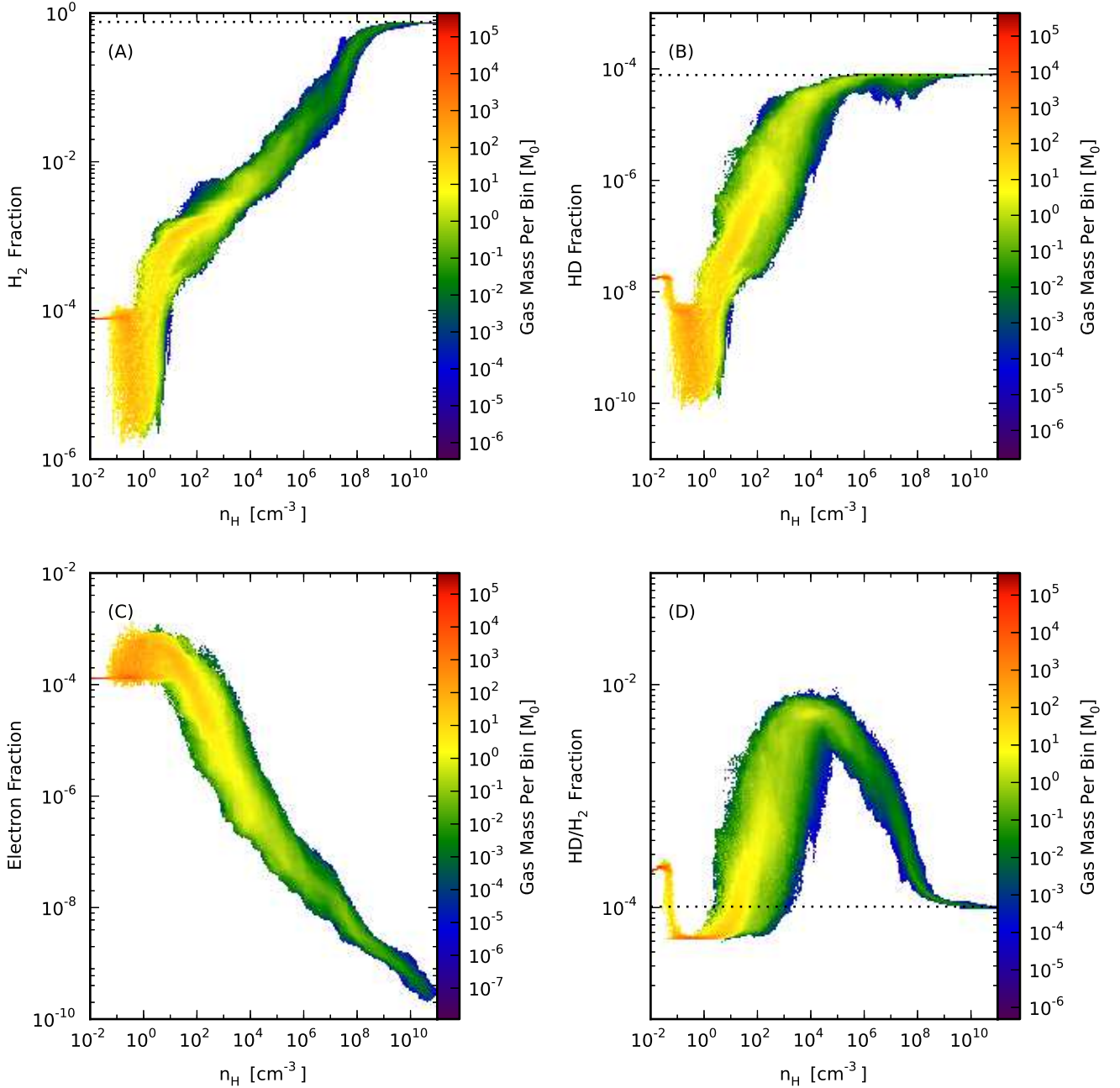}
   \centering
   \caption{The chemical state of the gas for the high mass fiducial model is shown. Panels A and B show the mass fractions of H$_2$ and HD respectively. Panel C shows the ionization fraction. Panel D shows the ratio of the HD to H$_2$ mass fractions. The dotted lines in Panels A and B show the mass fractions when H and D respectively are fully molecular, and the dotted line in Panel D shows the mass fraction ratio when both species are fully molecular.}
   \label{fig:chemical_evolution}
\end{figure*}

The physical evolution of the model may be understood by looking at the thermodynamic evolution of the gas, shown in Panel C of Figure \ref{fig:high_mass_physical_evolution}, and the chemical evolution, shown in Figure \ref{fig:chemical_evolution}. In low density regions, the gas is in free fall, and is heated by adiabatic compression. The relevant reaction rates are too slow to change the initial molecular chemistry, and radiative cooling is negligible compared to compressive heating.\par

At the accretion shock, the gas is rapidly heated to the virial temperature of the halo. In the high mass halo, the virial temperature is high enough that the gas enters the regime in which a small fraction of the H$_2$ and HD molecules are dissociated and some of the gas is ionized. As the gas becomes denser, the gas cools and the molecular fraction begins to increase.\par

The main coolants in the gas at are H$_2$, HD, and metals. At temperatures below 10,000 K, atomic hydrogen line cooling becomes negligible. H$_2$ is the most abundant species that is capable of radiative cooling, but is inefficient owing to the lack of a permanent dipole. Instead, the H$_2$ molecule must rely on rare quadrupole transitions between widely spaced energy levels, and by itself is unable to cool the gas below a temperature of around 200 K \citep{galli_1998}. HD, though rarer, has a permanent dipole moment and thus is able to cool more efficiently. Together, rotational transitions in HD and fine structure transitions in metals can effectively cool the gas to the CMB temperature floor. The ratio of HD and H$_2$ is set by the equilibrium rate of the reactions

\begin{align}
   \text{H}_2 + \text{D}^+ &\Rightarrow \text{HD} + \text{H}^+ \label{equation:H2_HD_reaction}\\ 
   \text{HD} + \text{H}^+ &\Rightarrow \text{H}_2 + \text{D}^+ \label{equation:HD_H2_reaction}
\end{align}

as described in \citet{omukai_2005}. Because of the differences in the energy levels of H$_2$ and HD, Equation \ref{equation:H2_HD_reaction} is preferred over Equation \ref{equation:HD_H2_reaction}, resulting in an HD/H$_2$ fraction that is higher than the overall D/H fraction by roughly 2 orders of magnitude \citep{galli_1998}. This fractionation is observed in Panel D of Figure \ref{fig:chemical_evolution}, which shows the HD/H$_2$ ratio. As the gas cools, the equilibrium abundance rapidly begins to favor HD production, which further increases cooling and in turn leads to more HD formation. For densities higher than $n_H \sim 10^5$ \cc, the deuterium is fully molecular.\par

The gas continues to collapse until either the temperature is too low to populate excited states in the coolants or the gas reaches the CMB temperature. For halos at $z=20$, we impose a CMB with temperature

\begin{equation} \label{equation:CMB_temp}
   T_{\text{CMB}} = 2.725~(1+z) = 57.225~\text{K}
\end{equation}

which enters into the heating equation for the gas and dust. The gas remains at the CMB temperature floor until a density of $n_H \sim 10^7~$ \cc is reached, at which point rapid formation of H$_2$ on dust grains briefly reheats the gas. At the highest densities, cooling via dust emission is able to efficiently lower the temperature of the gas, resulting in cooling in the higher-metallicity simulations.\par

\begin{figure*}
   \includegraphics[width=1.0\textwidth]{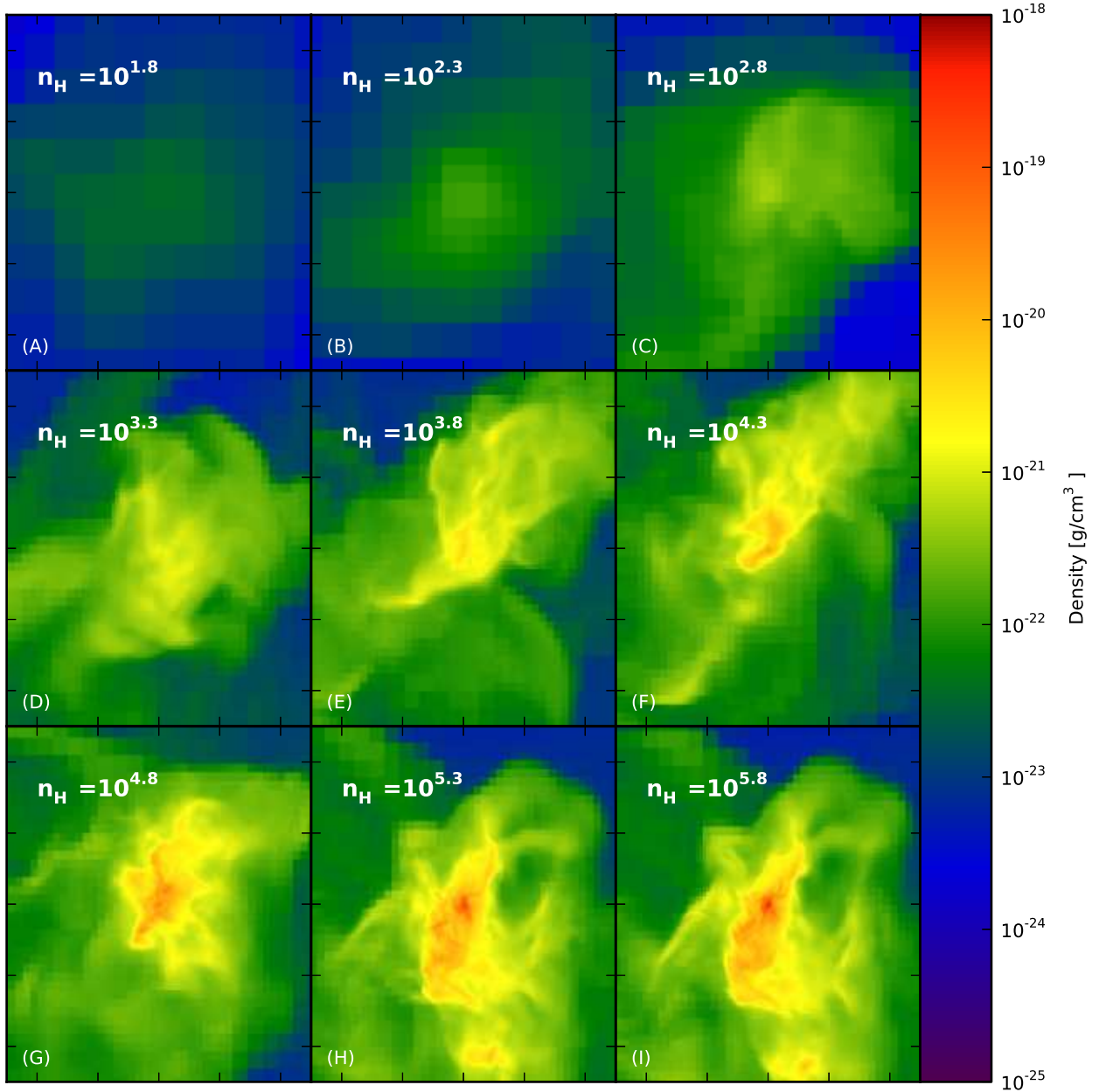}
   \centering
   \caption{Projections of average density through the densest point are shown as the central density increases. Each projection has a scale of 10pc. The gas is unstable to fragmentation whenever the Jeans mass decreases with increasing density, which occurs for densities between $n_H \sim 10^1$ and $n_H\sim 10^4$ \cc, but structure will only form when the central gas mass exceeds the local Jeans mass, which only occurs once the central density has increased above $n_H \sim 10^3$ \cc.}
   \label{fig:fiducial_evolution_thumbnails}
\end{figure*}

The formation of structure in the halo is governed by the thermodynamics of the gas during collapse. If the collapsing gas is able to cool with increasing density or if the temperature increases with density at a slower rate than $T \propto \rho^{1/2}$, the local Jeans mass will decrease. As the local Jeans mass sets the scale for fragmentation, the gas will be expected to fragment whenever the Jeans length is decreasing. Figure \ref{fig:fiducial_evolution_thumbnails} shows projections of density through the gas as the central density increases. At low densities, the mass of gas in the center region is below the local Jeans mass. As indicated in Panel C of Figure \ref{fig:high_mass_physical_evolution}, the gas is able to cool with increasing density for densities between $n_H \sim 10^1~$ \cc and $n_H \sim 10^5$ \cc. As the gas cools and density increases, the local Jeans mass is lowered below the central gas mass, causing perturbations to grow in the regime where the density is above $ n_H \sim 10^{2}$ \cc. \par

\section{Refinement Criteria} \label{section:varying_refinement_criteria}
To achieve the large dynamic range studied in our simulations, we selectively refine grid cells based on density, cooling time, and Jeans length, as described in Section \ref{section:refinement_criteria}. As part of this work, we have carried out a number of simulations wherein we vary the number of cells required to cover the Jeans length, $N_J$, in order to determine the minimum set of criteria needed to resolve the collapse.\par

As described in \citet{truelove_1998}, under-resolving the Jeans length in grid based codes can lead to artificial super-Jeans perturbations that may lead to spurious fragmentation. In tests of the collapse of a cloud with a Gaussian density profile, \citet{truelove_1998} concludes that the Jeans length should be covered by at least 4 cells at all times. However, this does not necessarily imply that the simulation is resolved enough to reveal pertinent details of fragmentation in the collapsing gas. Indeed, several studies (\citet{federrath_2011, turk_2012, latif_2013} and references therein) have found that at least 32-64 cells per Jeans length are necessary for resolving vorticity when modeling magnetic fields in Population III star formation.\par

\begin{figure}
   \includegraphics[width=0.5\textwidth]{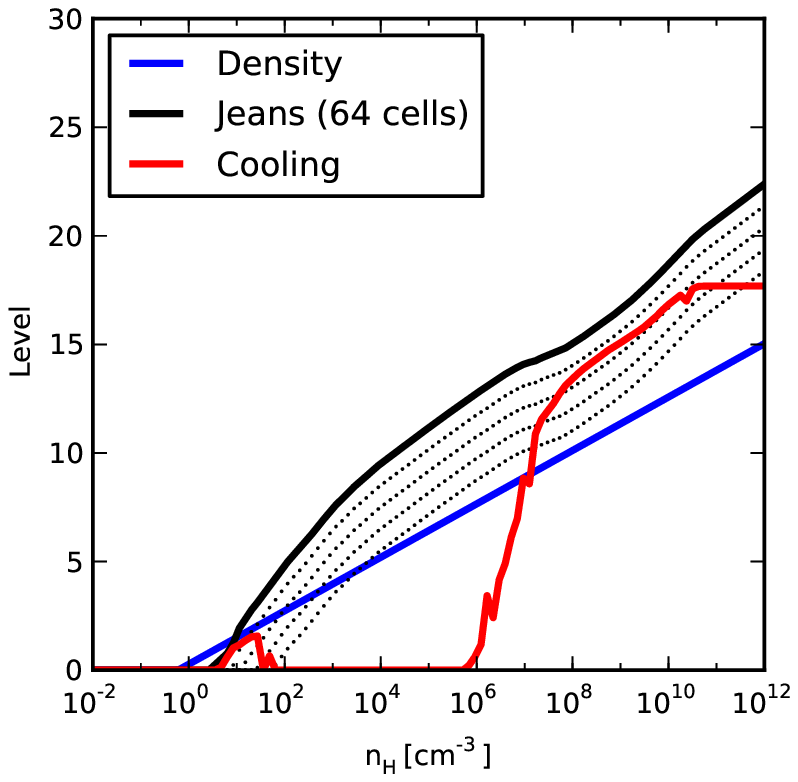}
   \centering
   \caption{The minimum refinement level for each refinement criterion is shown as a function of density for our high mass fiducial halo. In other words, each line represents the level to which the simulation would refine if only that criteria were applied. A cell will be refined until it is at the highest necessary refinement level, meaning that the actual level of refinement at a given density is indicated by the highest level in the plot above. The Jeans criterion and cooling time criterion are evaluated using the mass weighted average temperature and cooling time for each density. The solid black line shows the Jeans refinement level with 64 cells covering the Jeans length. From top to bottom, the dotted black lines show the level with 32, 16, 8, and 4 cells covering the Jeans length.}
   \label{fig:refinement_levels}
\end{figure}

To understand the effects of varying the strictness of the Jeans criterion on the physical phenomenon we are interested in, it is important to understand which refinement criteria dominate at different densities. In Figure \ref{fig:refinement_levels}, we show the minimum level to which a cell must be resolved as a function of density for each refinement criterion in our fiducial model. From Equation \ref{equation:mass_refinement}, it is easy to calculate the minimum grid level for which the mass refinement criteria is satisfied for a given density. To calculate the refinement level necessary to satisfy the Jeans and cooling criteria, which rely on the temperature and the cooling time in addition to the density, we use the average values of these quantities at each density from our fiducial model. Since a cell will be refined until all refinement criteria are met, the criterion with the largest minimum value will be the dominant criterion at a given density. In fact, if the required Jeans length coverage is set to $N_J=64$ or higher, the only place where the Jeans length will not be the dominant criterion is at the lowest densities, where fragmentation has not yet begun. From Figure \ref{fig:refinement_levels}, it can be seen that the Jeans refinement criterion is the dominant criterion at almost all densities. In addition, it is seen that for a range of densities, the Jeans criterion will be the dominant criterion even when the minimum number of cells covering the Jeans length is lowered. Thus, increasing the mandated Jeans length coverage will change the resolution over a wide range of densities and in general will increase the resolution of the simulation over a large range of spatial and mass scales as compared to the standard density-based criteria.\par

\begin{figure*}
   \includegraphics[width=1.0\textwidth]{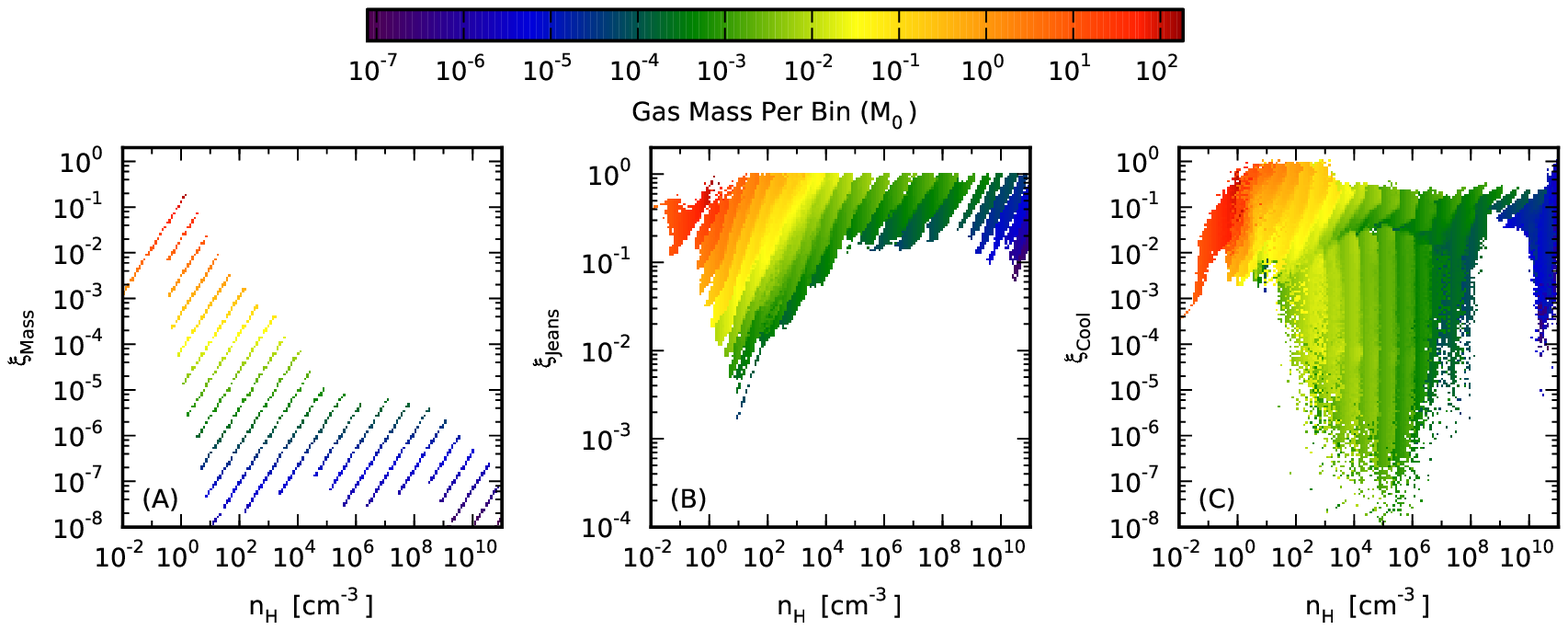}
   \centering
   \caption{The importance of different refinement criteria are shown for our $10^7 \Msun$ fiducial model. For each criteria, a cell is refined if one quantity (e.g. cell mass) is greater than a second quantity (e.g a minimum mass for refinement). The ratio of the two quantities are denoted by $\xi$, where $\xi_{\text{Mass}}$ is the ratio for mass refinement, $\xi_{\text{Jeans}}$ is the ratio for Jeans refinement, and $\xi_{\text{Cooling}}$ is the ratio for cooling based refinement. A cell should be flagged for refinement if $\xi$ for any criteria is greater than 1.0. At most densities, the Jeans criterion is the most dominant refinement criterion, with cooling time being important at low densities. Density based refinement is never important.}
   \label{fig:fiducial_refinement}
\end{figure*}

The lines shown in Figure \ref{fig:refinement_levels} are calculated using the mass-weighted average of the temperature and cooling time at a given density. While this approach is useful for finding the regimes when each criterion is dominant, it does not take into account variations in the temperature or cooling time of the gas at a given density, which may cause the minimum refinement level to vary. In particular, an average cooling time for gas near the CMB floor is not representative. Gas in that density regime with a temperature above the floor will cool, while gas with a temperature below the floor will heat, giving an average cooling time that is very long but ignoring that the actual cooling or heating time of the gas may be significantly shorter. In order to assess the importance of the different criteria on a cell-by-cell basis, we look at how close each cell in the simulation is to being refined. To do this, we evaluate the ratios

\begin{equation}
   \xi_{\text{Mass}} = \frac{M_{cell}}{M_{top} \times 8^{-0.3 \cdot l}}
\end{equation}

\begin{equation}
   \xi_{\text{Jeans}} = \frac{N_J \Delta x}{\lambda_J}
\end{equation}

\begin{equation}
   \xi_{\text{Cooling}} = \frac{t_{cool}}{t_{sound}}
\end{equation}

where $M_{cell}$ is the mass of the cell, $M_{top}$ is the mass of a top level grid cell, $N_J$ is the number of cells that must cover the local Jeans length, $\Delta x$ is the cell width, $\lambda_J$ is the local Jeans length, $t_{cool}$ is the cooling time, and $t_{sound}$ is the sound crossing time of a cell, $t_{sound} = \Delta x / c_s$. If any of these ratios are greater than 1, a cell will be refined. For our fiducial model, the distributions of $\xi_{\text{Mass}}$, $\xi_{\text{Jeans}}$, and $\xi_{\text{Cooling}}$ are shown in Figure \ref{fig:fiducial_refinement}. As expected, the density refinement criterion is not important for cells with densities greater than $n_H \sim 10^1$ \cc. At low densities, both the Jeans and cooling time criteria are close to being met in a large number of cells. At higher densities, only the Jeans criterion is close to being met, indicating that it is indeed the only important refinement criterion. Panel C of Figure \ref{fig:fiducial_refinement}, however, does indicate that the cooling time refinement is likely to be dominant at low densities for some cells.\par

Further tests of our fiducial model in which only the Jeans criterion is used show similar overall evolution to the runs with all three refinement criteria, but for the high mass run there are a small number of cells that are not refined, but ordinarily would meet the cooling criterion for refinement. Thus, the cooling time criterion is necessary in some circumstances to fully resolve the collapse of the gas. For our low mass model, we find that the Jeans criteria is always dominant because temperatures are lower and thus the cooling time is longer than in the high mass case.\par

\begin{figure*}
   \includegraphics[width=1.0\textwidth]{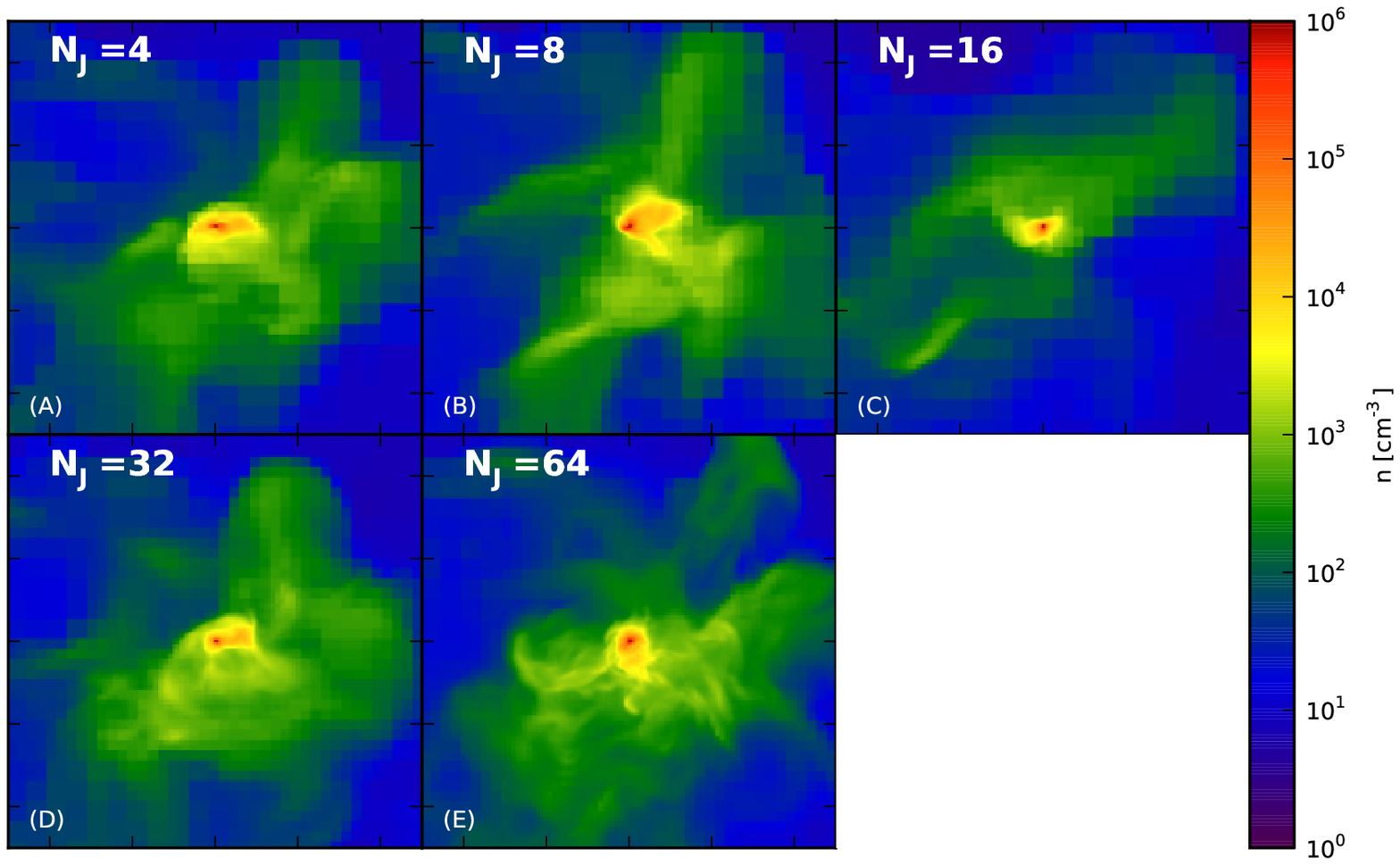}
   \centering
   \caption{Projections of average density through the densest point in the simulation for runs with different Jeans refinement criteria for our $10^7 \Msun$ fiducial halo. Each projection has a width of 10 pc, and is taken when the central density has reached $n_H=10^{10}$ \cc.}
   \label{fig:hm_jeans_comparison_10pc}
\end{figure*}

\begin{figure*}
   \includegraphics[width=1.0\textwidth]{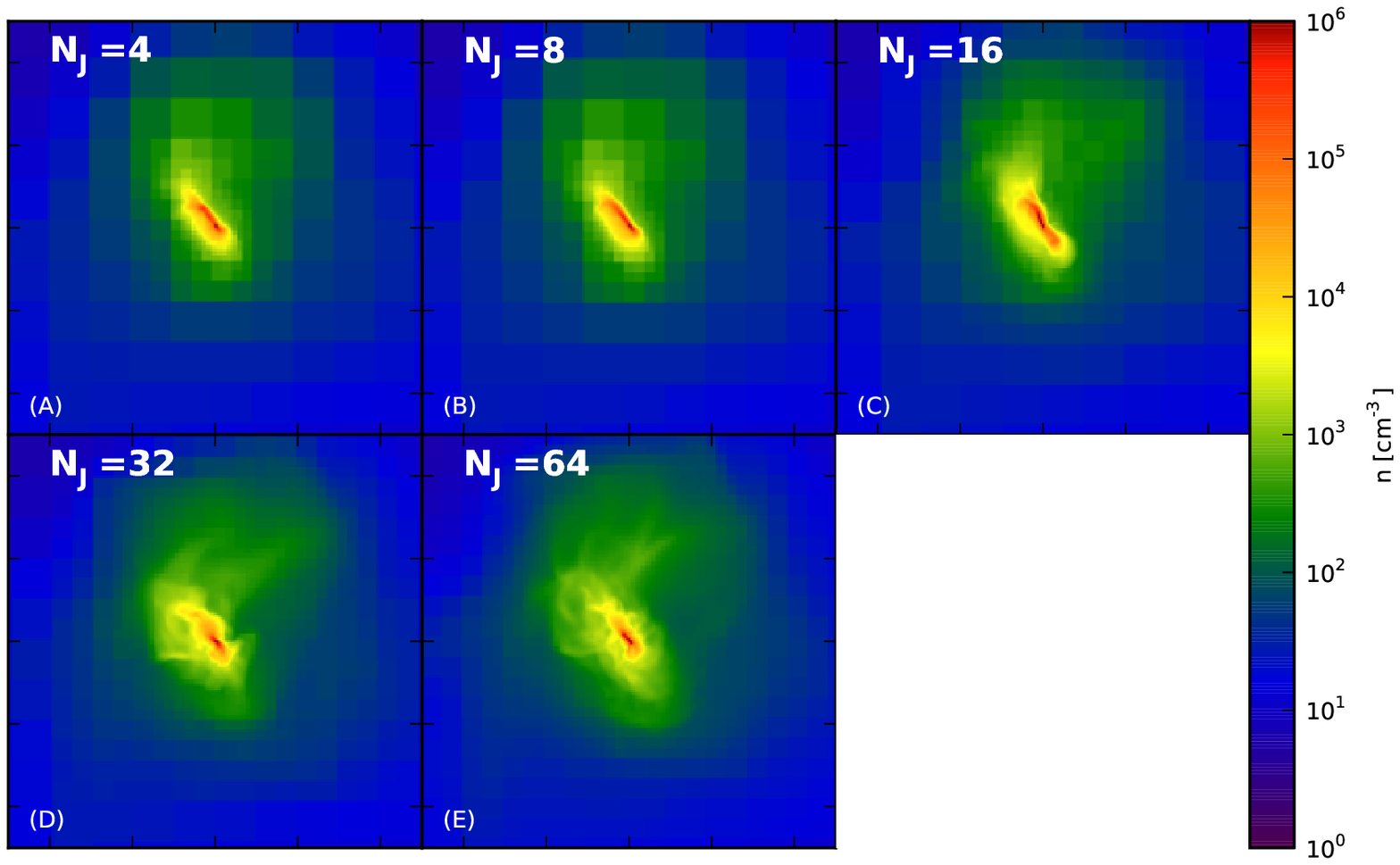}
   \centering
   \caption{Projections of average density through the densest point in the simulation for runs with different Jeans refinement criteria for our $10^6 \Msun$ fiducial halo. Each projection has a width of 10 pc, and is taken when the central density has reached $n_H=10^{10}$ \cc.}
   \label{fig:lm_jeans_comparison_10pc}
\end{figure*}

Having established that the Jeans criteria is nearly always the dominant factor in setting grid resolution, the question becomes how strict our refinement criteria needs to be in order to properly resolve the collapse and fragmentation of the cloud. To understand the effects of resolution criteria, we have carried out a series of runs (described in Table \ref{table:jeans_table}) where we vary the number of cells that must cover the Jeans length from the Truelove criterion of $N_J=4$ to a maximum of $N_J=64$, the limit of what is computationally feasible for our study. The final state of this suite of simulations is shown at a scale of 10 pc in Figure \ref{fig:hm_jeans_comparison_10pc} for our high mass halo and in Figure \ref{fig:lm_jeans_comparison_10pc} for our low mass halo.\par

From these projections, it is clear that the evolution of the gas is affected by the level of resolution, even when the Truelove criterion is substantially exceeded. The runs with $N_J=32$ and $N_J=64$ show fragmentation on small scales that is not present in the other runs. The differences stem from increased resolution of the gas at the densities where fragmentation occurs, which leads to an increase in the strength of perturbations with large wave number. From the projections, it is clear that a ``phase transition'' of sorts occurs between $N_J=16$ and $N_J=32$ cells, but there are also hints of additional fragmentation at $N_J=64$ cells. We note that using $N_J=128$ when simulating the high mass halo caused a sharp increase in the number of grid cells in the simulation, and the run was terminated when it was determined to be computationally infeasible. For the rest of the runs in this study, we choose to use 64 cells to cover the Jeans length in order to resolve small scale perturbations while maintaining computational feasibility, but caution the reader that further increasing the resolution may have non-negligible effects.\par

\begin{figure*}
   \includegraphics[width=1.0\textwidth]{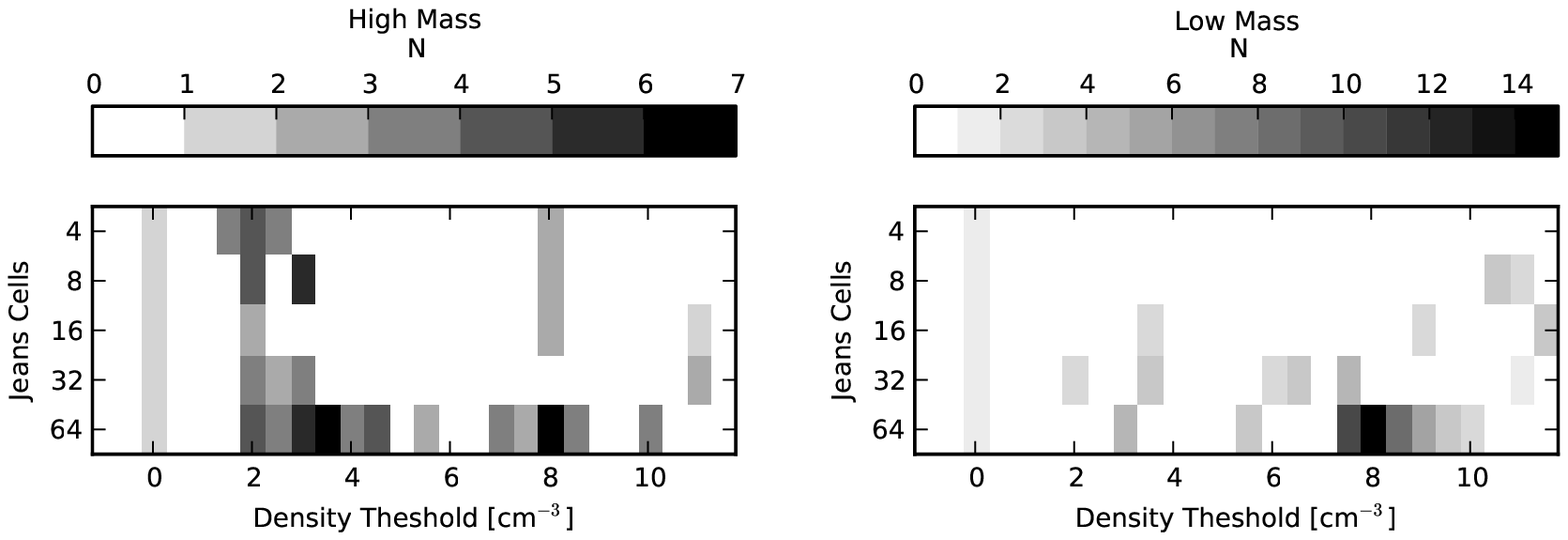}
   \centering
   \caption{The number of gravitationally bound or nearly bound clumps is shown for runs with with different levels of Jeans refinement. The y-axis shows the number of cells that must cover the local Jeans length at all times. We identify clumps using a contouring algorithm, and keep only those clumps that are close to being gravitationally bound and which will become bound if cooling continues. The number of clumps in each half dex contour interval is shown above. Clump finding is performed when each run reaches a central density of $10^{10}$ \cc. For a full description of the clump finding routine, see Section \ref{section:clump_finding}.}
   \label{fig:jeans_clumps}
\end{figure*}

In order to quantify the effect of Jeans resolution on fragmentation, we use the clump finding algorithm described in Section \ref{section:clump_finding} to find the number of potentially bound clumps in each simulation, which is shown in Figure \ref{fig:jeans_clumps}. We observe a trend of increasing fragmentation (inferred from the increase in the number of identified clumps) with higher resolution of the Jeans length. The increase in fragmentation is seen at all densities, and is particularly evident at higher densities ($n_H > 10^3$ \cc). The low mass halo shows less fragmentation overall, but the trend of increasing fragmentation with increasing strictness of the Jeans resolution criteria holds.\par

\section{Effects of Physical Parameter Variation on Gas Fragmentation}\label{section:exploration_of_parameter_space}
\subsection{Metallicity}\label{section:varying_metallicity}
Several studies have found that the introduction of metals has a strong effect on the cooling properties of star forming gas. As the fraction of metals increases, the gas is able to cool more efficiently. This in turn may lead to increased fragmentation, and is the reason that increasing metallicity has been proposed as the driving factor behind the purported transition between a high characteristic mass Population III IMF and a lower characteristic mass metal-enriched IMF. To understand the effects of metallicity in our model, we perform a series of runs (see Table \ref{table:metallicity_table}) where the metal content of the gas is varied from a uniform metallicity of $Z=0$ (metal free, primordial gas) to $Z=10^{-2}~\Zsun$. In these simulations, we assume that a fixed fraction ($9.23 \times 10^{-3}$ by mass) of the metals are in the form of dust.\par

\begin{figure*}
   \includegraphics[width=1.0\textwidth]{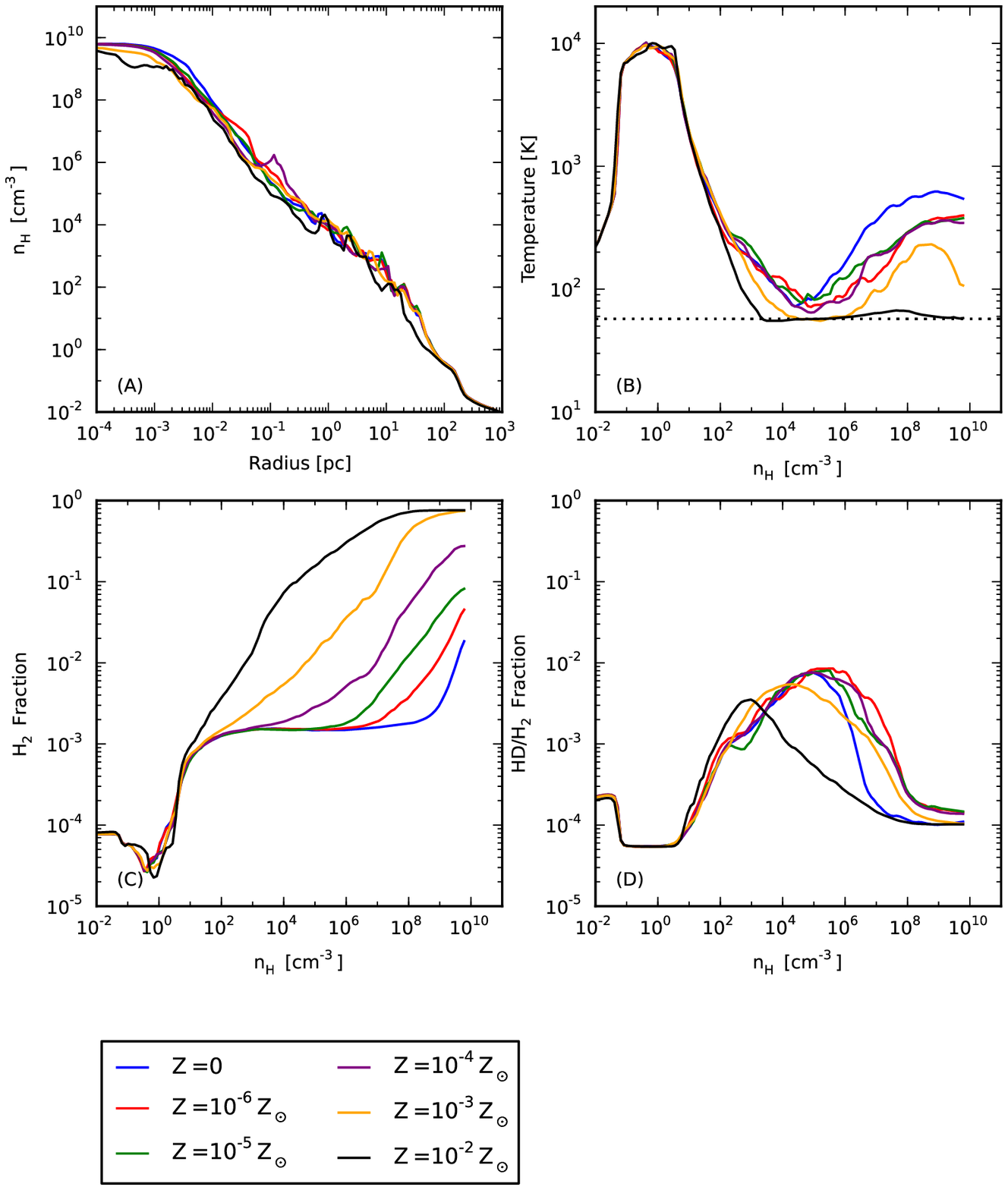}
   \centering
   \caption{Comparison of the physical, thermal, and chemical state of our high mass halo as metallicity is varied, at the point when the simulation reaches a central density of $n_H=10^{10}~$ \cc. Panel A shows spherically averaged, mass weighted density as a function of radius. Runs with higher metallicity collapse faster, leading to lower densities in the regions surrounding the densest point. Panel B shows spherically averaged, mass weighted temperature as a function of density. As metallicity is increased, the gas is able to cool to lower temperatures. The CMB temperature is indicated by a dashed line. Panel C shows the molecular hydrogen mass fraction as a function of density. In the metal free case, molecular hydrogen is formed primarily through the three body process, which does not become effective until densities of $n_H \sim 10^8$ \cc. As metallicity is increased, dust catalyzed reactions become the dominant mode of H$_2$ formation. Panel D shows the ratio of the HD to H$_2$ mass fractions, which is enhanced over the atomic value through chemical fractionation.}
   \label{fig:metallicity_comparison_profiles_high_mass}
\end{figure*}

\begin{figure*}
   \includegraphics[width=1.0\textwidth]{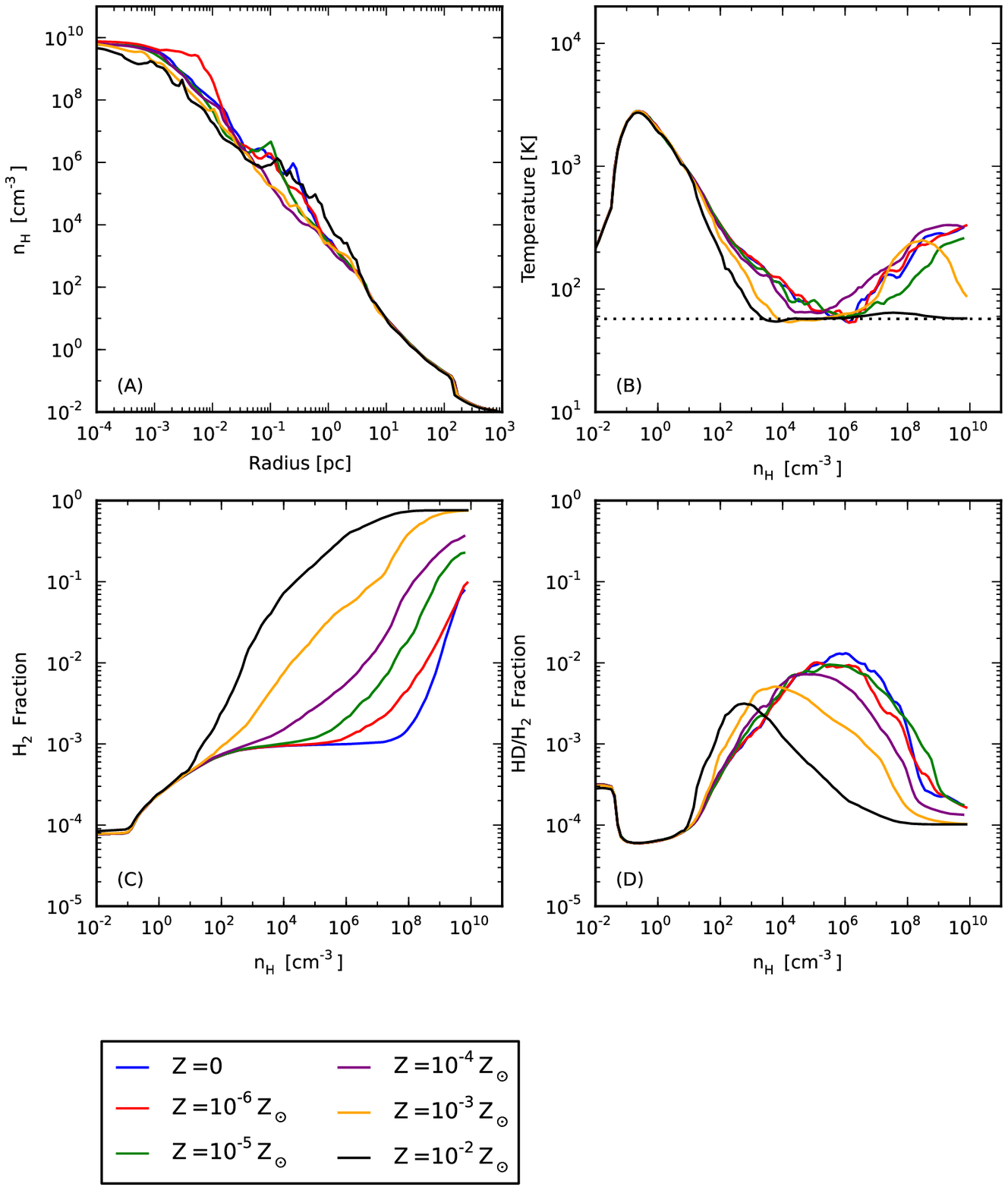}
   \centering
   \caption{Same as Figure \ref{fig:metallicity_comparison_profiles_high_mass}, but for the low mass halo.}
   \label{fig:metallicity_comparison_profiles_low_mass}
\end{figure*}

The effect of varying metallicity on the physical and thermodynamic evolution of the gas is shown in Figure \ref{fig:metallicity_comparison_profiles_high_mass} for the high mass halo and in Figure \ref{fig:metallicity_comparison_profiles_low_mass} for the low mass halo. Increasing the amount of metals alters the cooling rate in three ways. First, metals directly cool the gas, allowing the temperature to reach the CMB floor more quickly and at lower densities. Second, increasing the metallicity increases the amount of dust present. As dust-mediated reactions become the dominant molecular formation channel, H$_2$ and HD are able to form efficiently at densities below $n_H=10^8$ \cc, when 3-body reactions become effective. This leads to more cooling at lower densities. Thirdly, dust itself becomes an effective coolant at densities above $n_H=10^9$ \cc.\par

\begin{figure*}
   \includegraphics[width=1.0\textwidth]{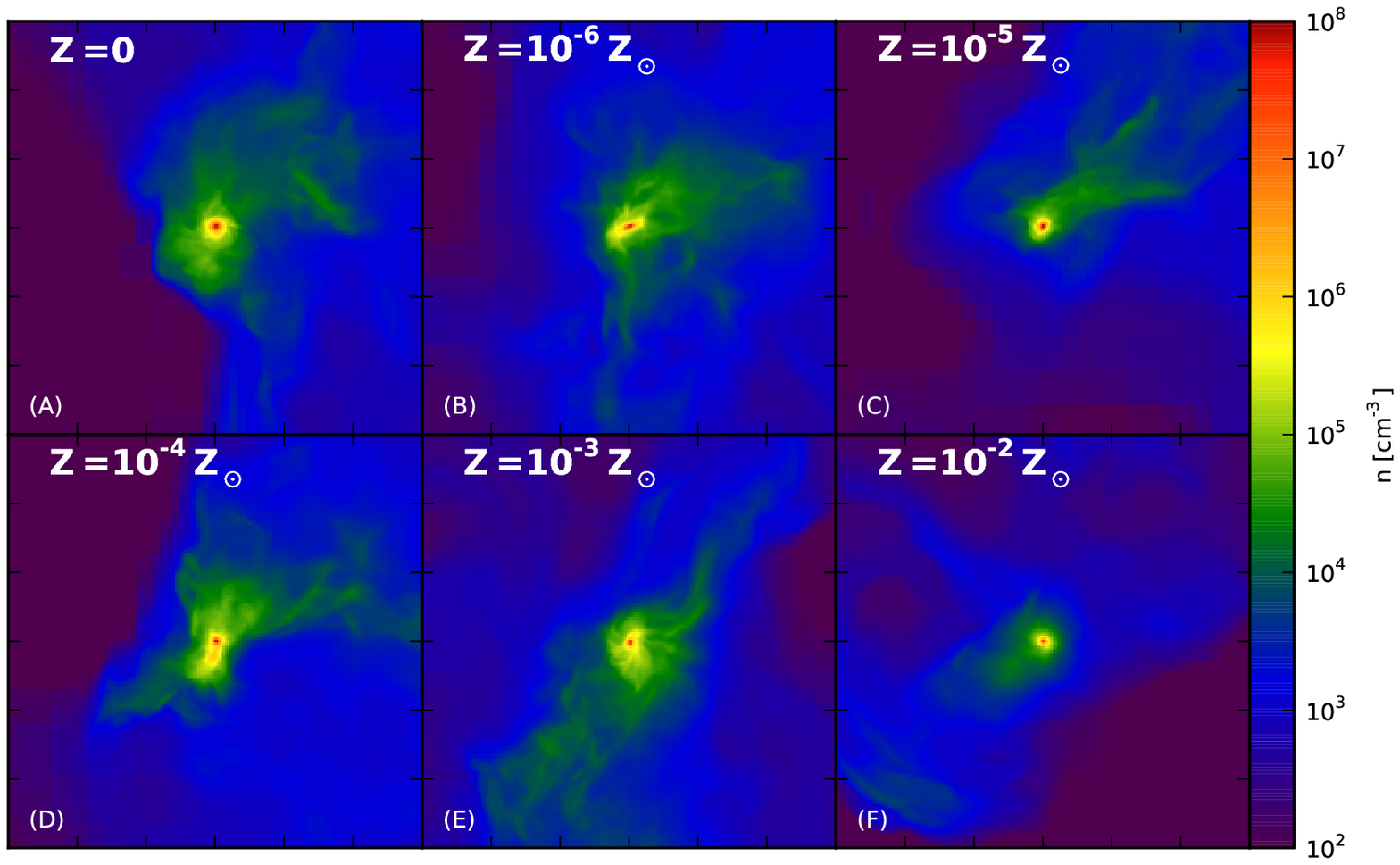}
   \centering
   \caption{Projections of average density through the densest point in the simulation for runs with different metallicity for our $10^7 \Msun$ halo. Each projection has a width of 3 pc, and is taken when the central density has reached $n_H=10^{10}$ \cc.}
   \label{fig:hm_metallicity_comparison_3pc}
\end{figure*}

\begin{figure*}
   \includegraphics[width=1.0\textwidth]{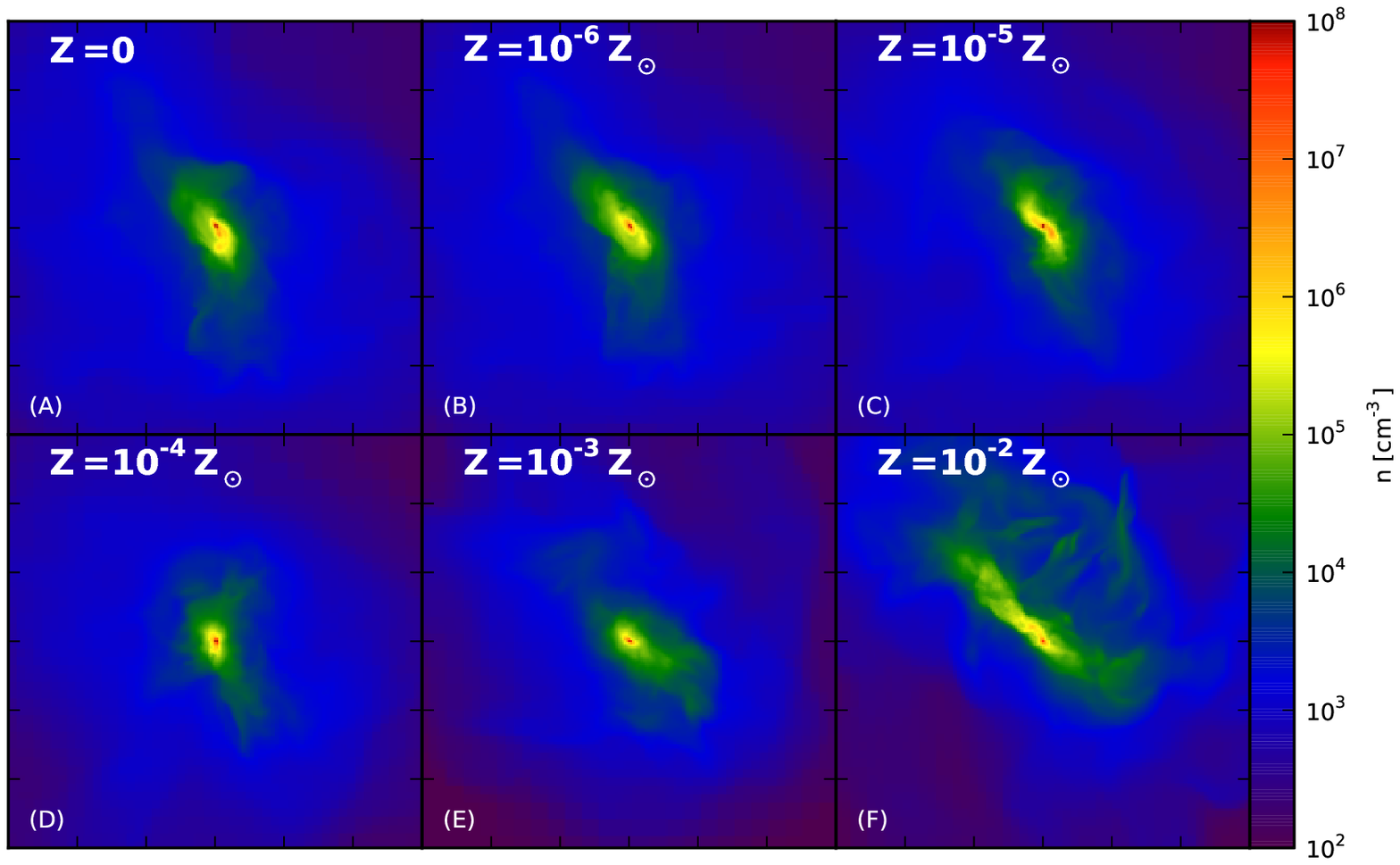}
   \centering
   \caption{Projections of average density through the densest point in the simulation for runs with different metallicity for our $10^6 \Msun$ halo. Each projection has a width of 3 pc, and is taken when the central density has reached $10^{10}$ cm$^{-3}$.}
   \label{fig:lm_metallicity_comparison_3pc}
\end{figure*}

The effects of metals on the physical state of the gas may be understood by looking at projections of gas density in the core of the halo. In Figure \ref{fig:hm_metallicity_comparison_3pc}, we show the final state of the high mass halo at a scale of 3 pc. In Figure \ref{fig:lm_metallicity_comparison_3pc}, we show the same plots for the low mass halo.\par

At high metallicities, the gas is able to cool rapidly, meaning that the densest region will be able to collapse before a large mass of gas has built up in the core. For low metallicity and metal-free gas, however, the gas must rely on H$_2$ and HD to cool. The collapse is delayed, which gives the core more time to grow, leading to a larger mass of dense gas in the central region. This is clearly seen in Figure \ref{fig:hm_metallicity_comparison_3pc} (particularly in Panel A), where the densest regions in the low metallicity runs are surrounded by more gas than in the high metallicity runs. The accretion rate is lower in the low mass halo, and the trend is not as clear. From the collapse times given in Table \ref{table:metallicity_table}, it is evident that there is a trend of faster collapse with increasing metallicity.\par

\begin{figure*}
   \includegraphics[width=1.0\textwidth]{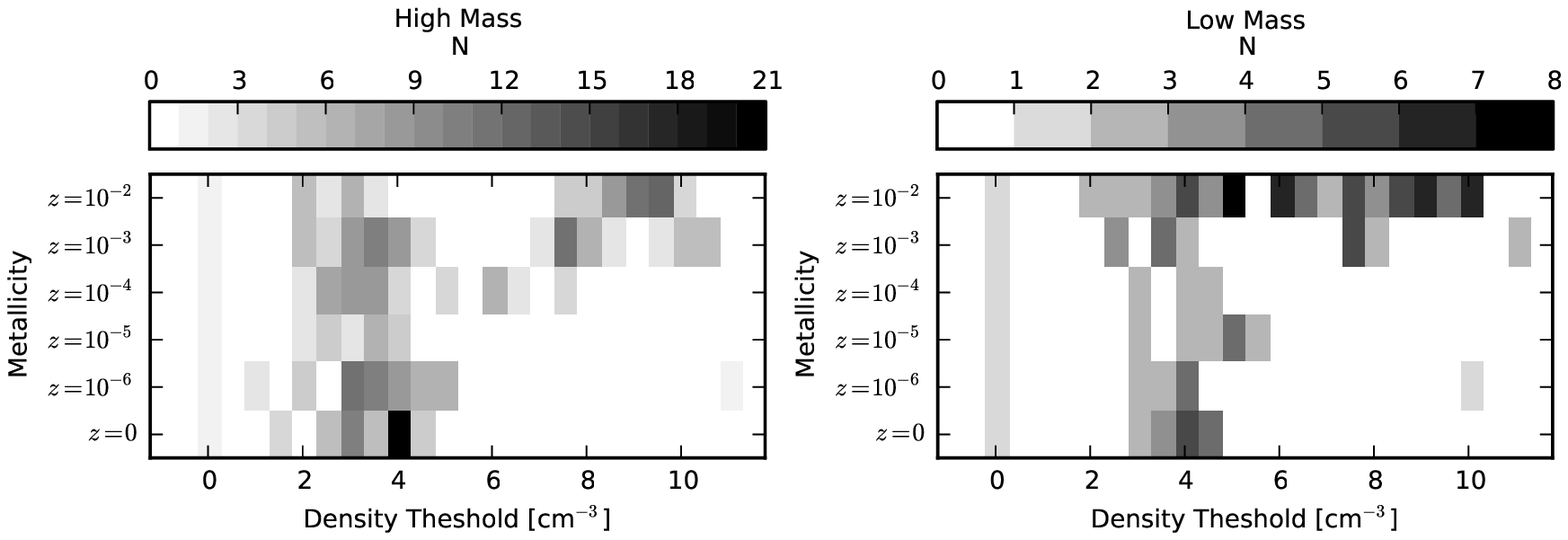}
   \centering
   \caption{The number of bound or potentially bound clumps identified by our clump finding algorithm as partially bound for runs in which metallicity is varied.}
   \label{fig:metallicity_clumps}
\end{figure*}

We expect that as the cooling rate increases, the amount of fragmentation will increase. In Figure \ref{fig:metallicity_clumps}, we show the distribution of clumps as a function of $n_H$ for the high and low mass halos. In the density range $10^1 < n_H < 10^5$ \cc, all runs show evidence of fragmentation, with slightly more fragmentation at higher metallicities. Here, the gas is able to fragment because the temperature is decreasing with increasing density. After the gas reaches the CMB floor it is no longer able to cool as density increases, inhibiting further fragmentation. As H$_2$ is formed, thermal energy is injected into the gas. The halos with higher metallicity are able to effectively radiate away this energy, which allows for more fragmentation at higher densities. In our high mass halo, only the simulations with metallicities about $10^{-3}~\Zsun$ show evidence of fragmentation at $n_H \gtrsim 10^7$ \cc. We note that cooling from dust may lead to further fragmentation in the lower metallicity runs at higher densities. As noted by \citet{schneider_2006} and \citet{schneider_2010}, dust cooling may lead to fragmentation in haloes with metallicities above $10^{-6}~\Zsun$ at densities above $n_H=10^{12}$, where the gas and dust temperatures couple.\par 

\begin{figure*}
   \includegraphics[width=1.0\textwidth]{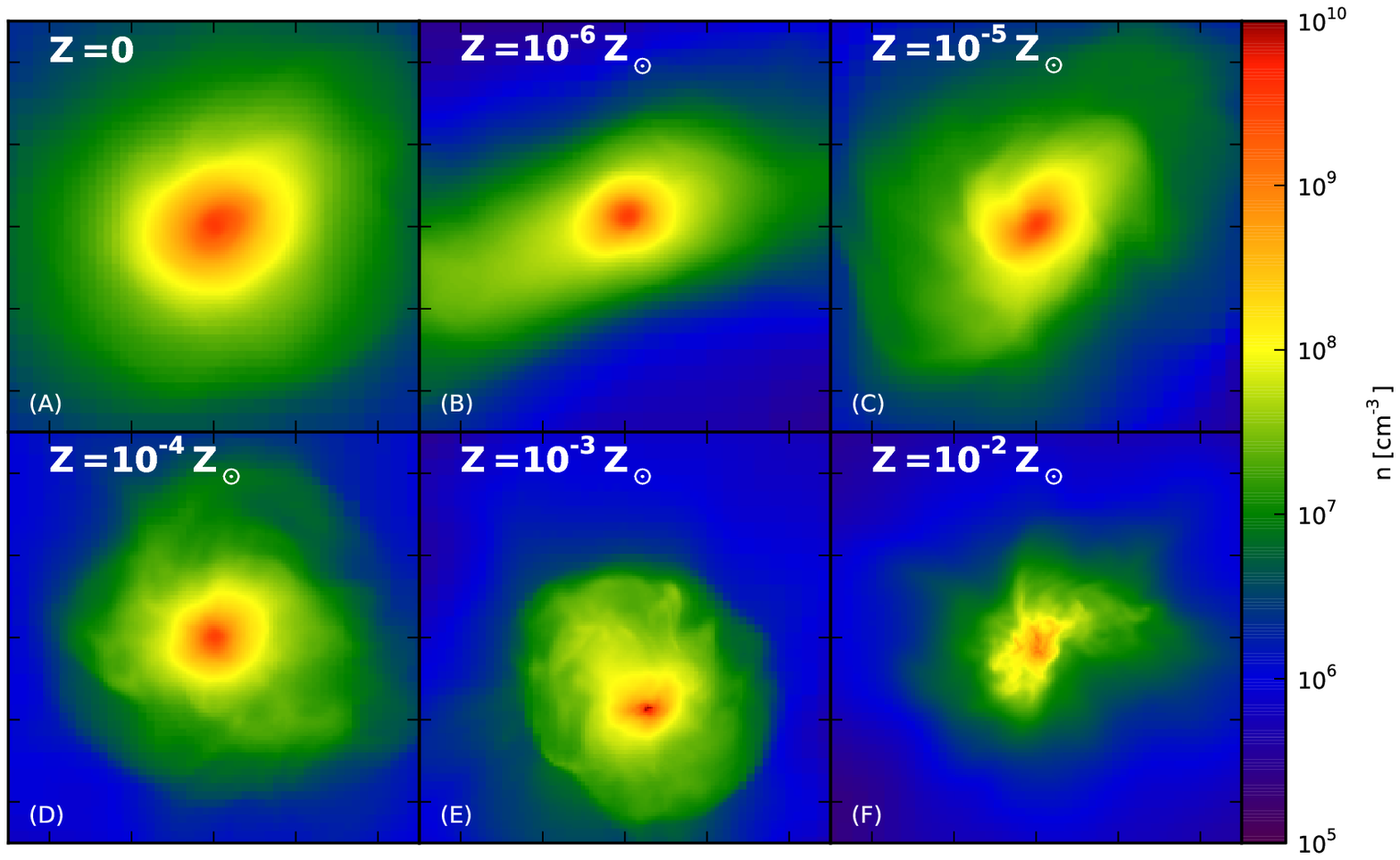}
   \centering
   \caption{Projections of average density through the densest point in the simulation for runs with different metallicity for our $10^7 \Msun$ halo at the scales where the clump finder finds evidence of fragmentation in high mass halos. Each projection has a width of 0.05 pc, and is taken when the central density has reached $10^{10}$ cm$^{-3}$. Substructure is evident in the higher metallicity runs (shown in the bottom row), and the clump finder confirms that some of these structures may become gravitationally bound if cooling persists.}
   \label{fig:hm_metallicity_comparison_005pc}
\end{figure*}

For those simulations where the clump finder indicates fragmentation at higher densities, it is interesting to see what effect cooling is having on the gas structure at the relevant scales. Figure \ref{fig:hm_metallicity_comparison_005pc} shows projections through the core at a width of 0.05 pc, which encompasses the density range above $n_H \sim 10^6$ \cc, at the density regime the clump finder indicates that metallicity affects fragmentation. The projections indicate that the gas in the core does indeed form a whispy substructure for high metallicity gas. Although the density contrast is small, the clump finder indicates that some of these structures are marginally gravitationally bound, giving rise to the possibility that some of these structures could become protostars.\par

\subsection{Spin} \label{section:varying_spin}
We have performed a series of runs where we vary the initial magnitude of angular momentum in our halos, as described by the spin parameter $\lambda$ (Equation \ref{equation:spin_parameter} in Section \ref{section:initial_velocity_profile}). For our high and low mass halos, we vary the spin parameter from $\lambda=0.0$ to $\lambda=0.1$, spanning the likely range of spin parameters for cosmological halos (e.g. \citet{yoshida_2003}). The turbulent component of the velocity also imparts some angular momentum to the gas, and is kept constant throughout these simulations. As the turbulent field is isotropic and the largest length scale is substantially smaller than the virial radius of the halo, the turbulent field should provide no net rotation. The full list of spin parameter-related simulations can be found in Table \ref{table:spin_table}.\par

\begin{figure*}
   \includegraphics[width=1.0\textwidth]{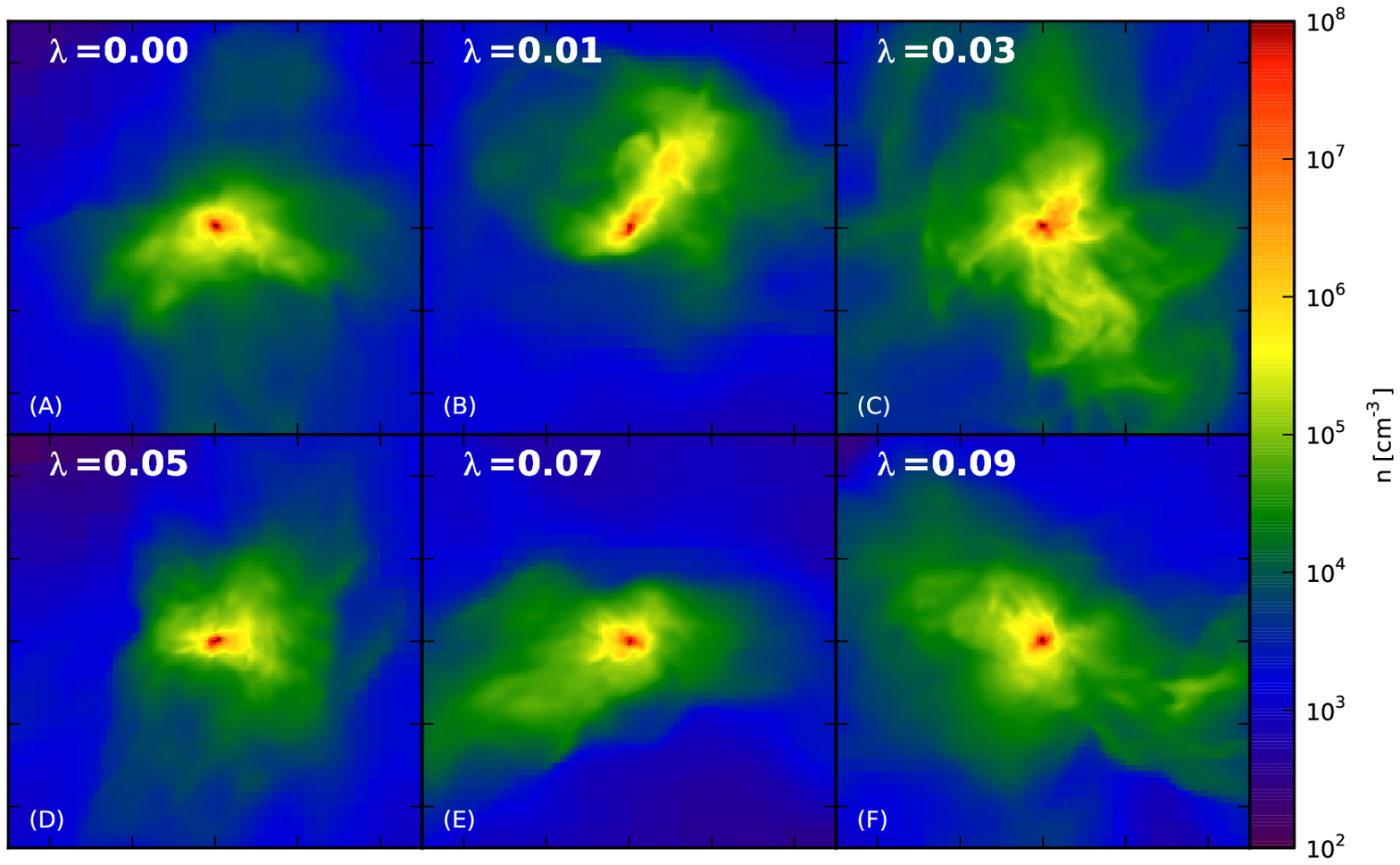}
   \centering
   \caption{Projections through the core of our simulations in which the spin parameter is varied. Although differences in structure are observed, there is no systematic trend in the fragmentation with increasing spin. The width of each image is 3.0 pc.}
   \label{fig:hm_spin_comparison_3pc}
\end{figure*}

\begin{figure*}
   \includegraphics[width=1.0\textwidth]{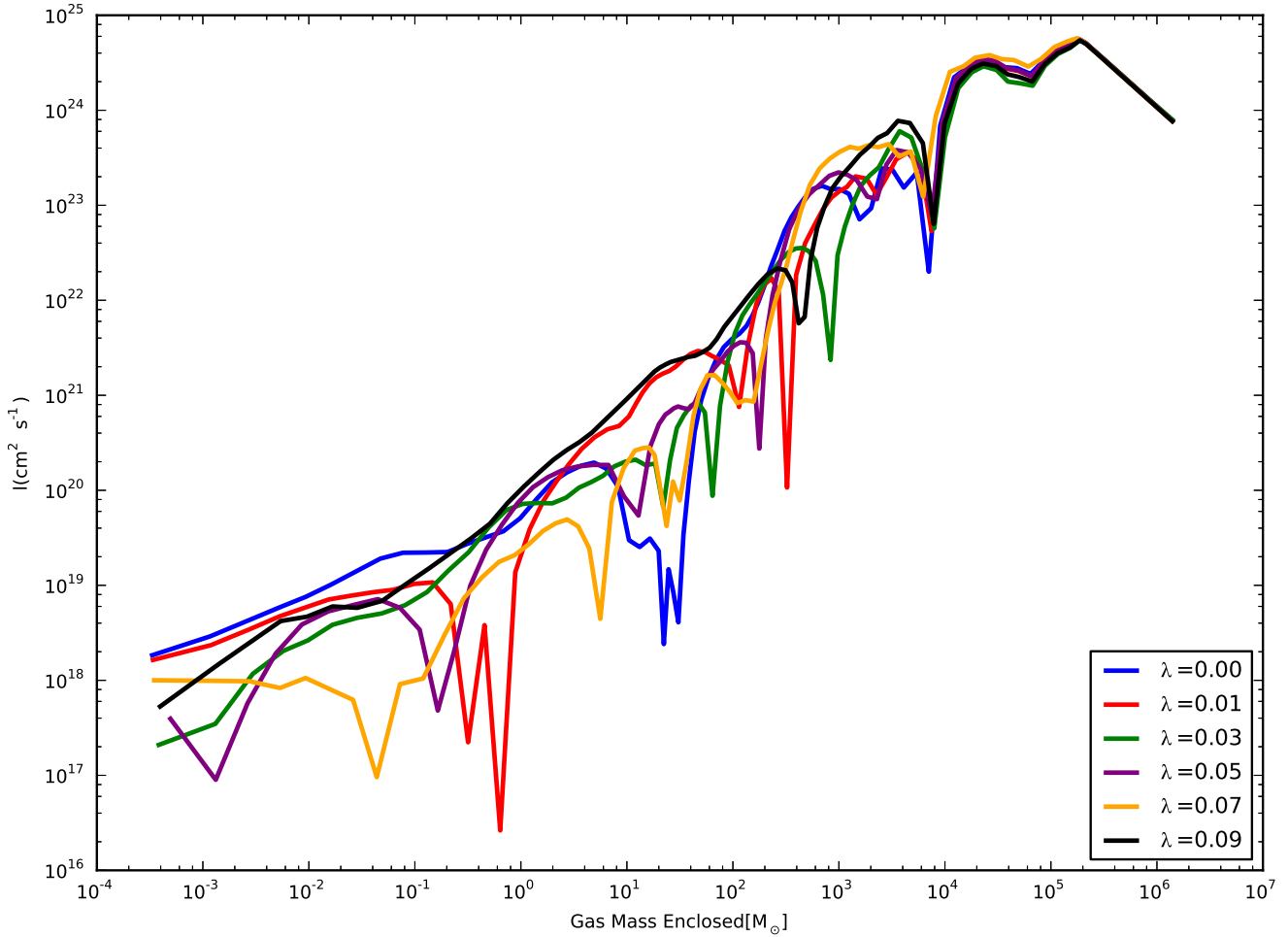}
   \centering
   \caption{Average angular momentum vs. mass enclosed for runs with different initial spin parameters for our high mass halos when the central density of each simulation reaches $n_H=10^{10}$ \cc. By this point, turbulence has randomized the distribution of angular momentum.}
   \label{fig:hm_spin_angular_momentum}
\end{figure*}

\begin{figure*}
   \includegraphics[width=1.0\textwidth]{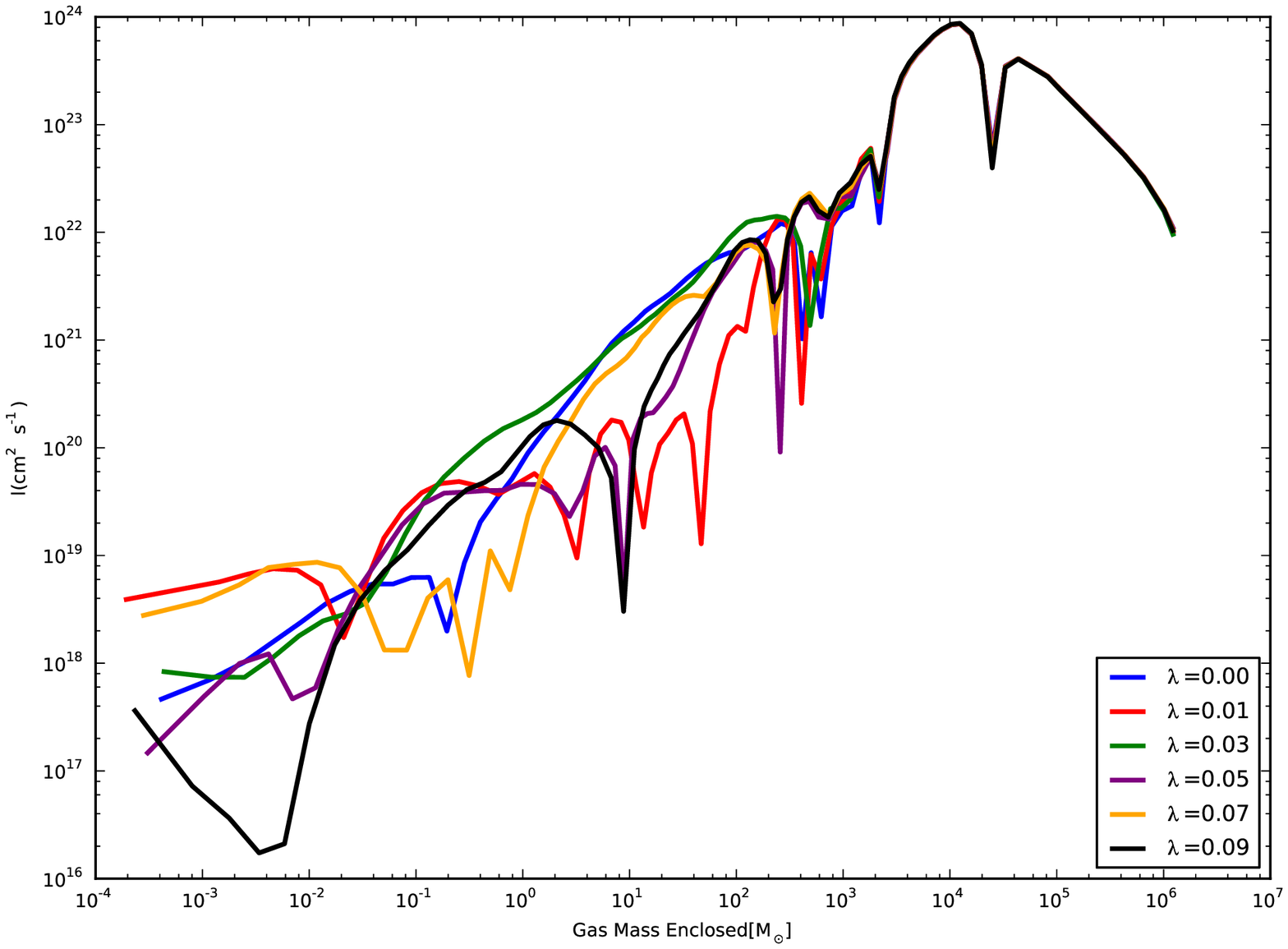}
   \centering
   \caption{Same as Figure \ref{fig:hm_spin_angular_momentum}, but for the low mass halo.}
   \label{fig:lm_spin_angular_momentum}
\end{figure*}

Projections of the cores of simulations with different spin parameters in the high mass halo are shown in Figure \ref{fig:hm_spin_comparison_3pc} at a scale of 3  pc. While there are substantial differences between the different runs, no clear trend relating fragmentation or gas density profile to initial spin emerges. This result is easy to understand in the context of Figures \ref{fig:hm_spin_angular_momentum} and \ref{fig:lm_spin_angular_momentum}, in which we plot average angular momentum as a function of mass enclosed for the high and low mass halos respectively. Turbulence is able to efficiently transport angular momentum away from the inner regions of the halo and normalize the angular momentum distribution for all runs to roughly the same level, such that the dynamics of the core are dominated by the turbulent motion rather than by initial rotation. The differences we observe are more likely due to stochastic effects resulting from the perturbation of the initial conditions.\par

\begin{figure*}
   \includegraphics[width=1.0\textwidth]{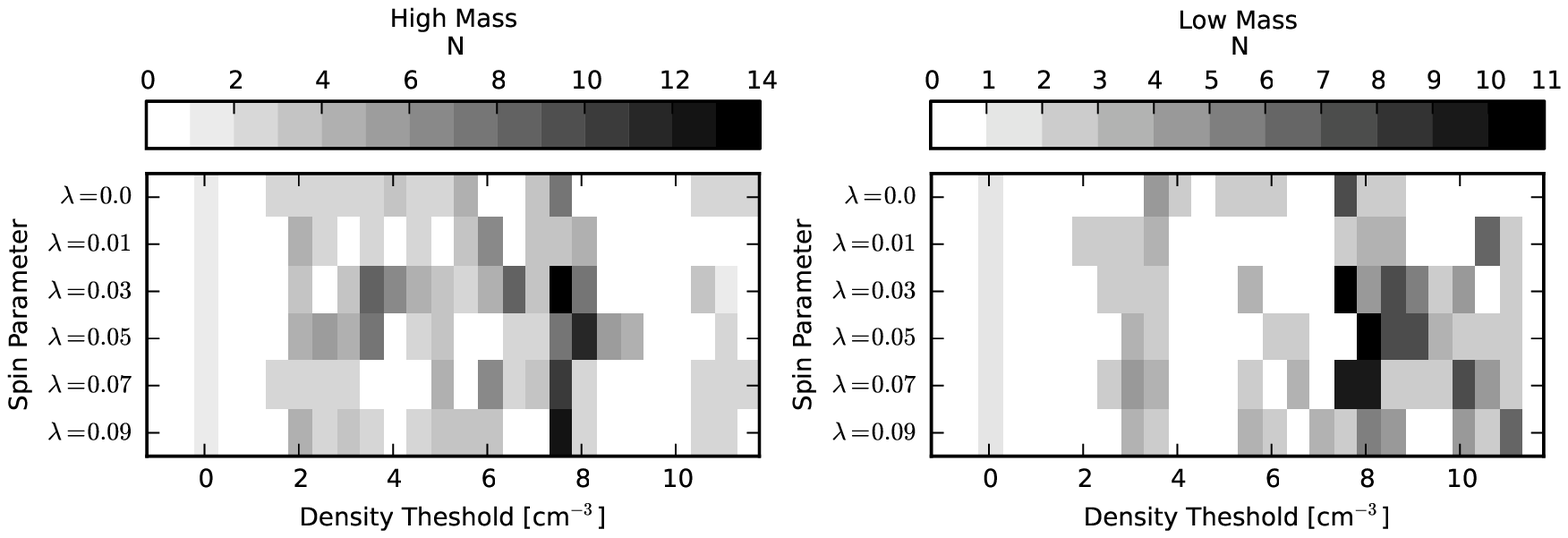}
   \centering
   \caption{The number of bound or nearly bound clumps is shown for runs with with different initial spin parameter as identified by our clump finder. The number of clumps in each half dex contour interval is shown above. Clump finding is performed when each run reaches a central density of $n_H=10^{10}$ \cc. The fragmentation profile confirms that spin has no clear effect on fragmentation beyond perturbing the initial conditions.}
   \label{fig:spin_clumps}
\end{figure*}

In order to confirm that the initial spin does not correlate with the amount of fragmentation in the halo, we identified clumps in each simulation using the mechanism as described in the previous sections. The number of clumps in each half-dex bin, shown in Figure \ref{fig:spin_clumps}, show no identifiable relationship between fragmentation and spin parameter.\par

\subsection{Turbulence} \label{section:varying_turbulence}
\begin{figure*}
   \includegraphics[width=1.0\textwidth]{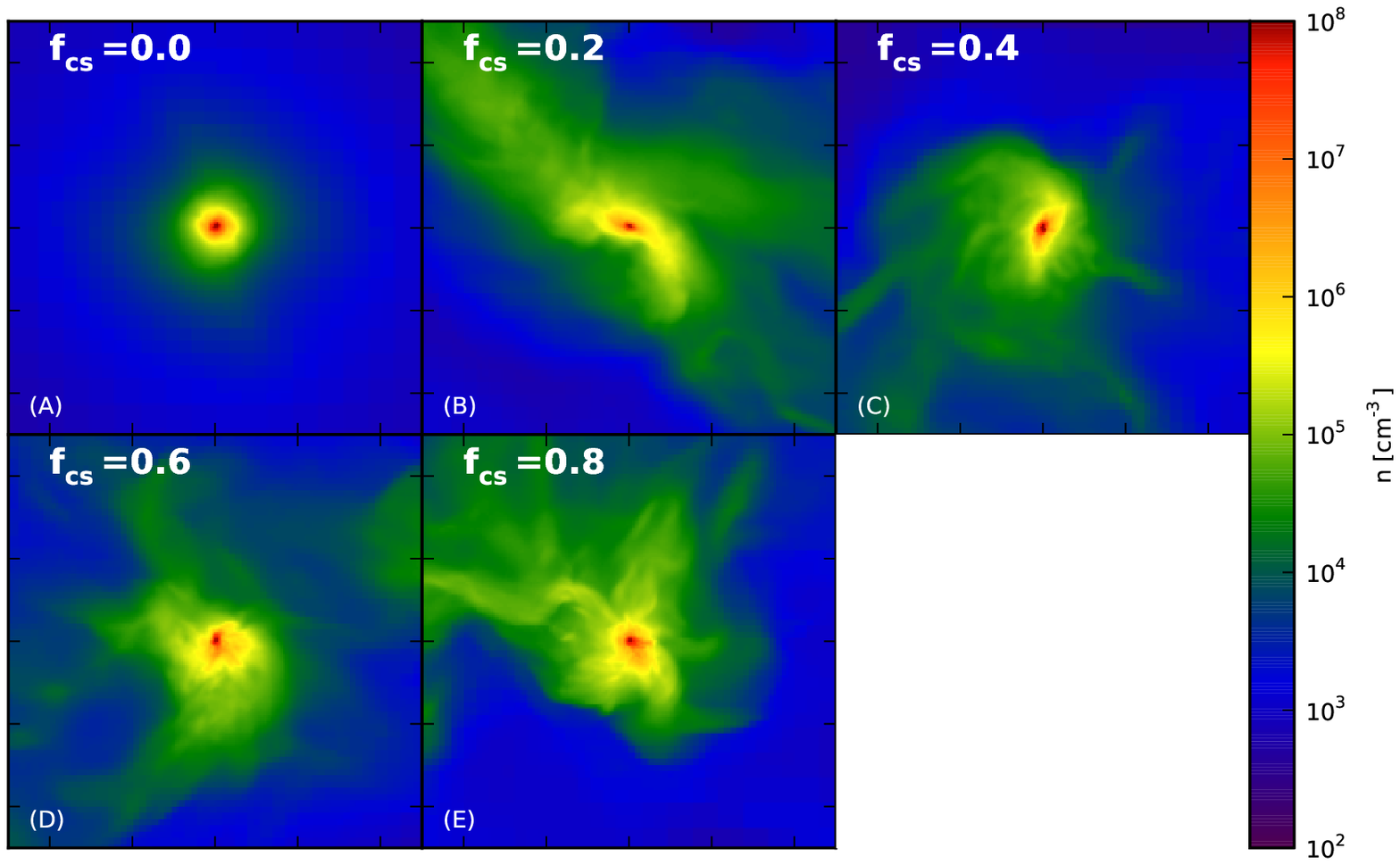}
   \centering
   \caption{Projections through the core of our simulations in which the amount of initial turbulence is varied. The scale of each image is 3.0 pc.}
   \label{fig:hm_turbulence_comparison_3pc}
\end{figure*}

To study the effects of the level of turbulent motion on the evolution of our model, we simulated our high and low mass halos with varying levels of turbulence. In our simulations, the RMS velocity of the initial turbulent field is normalized to some fraction $f_{cs}$ of the virial sound speed of the halo. Here, we show results for halos with values of $f_{cs}$ from 0.0 (no turbulence) to 0.8 (trans-sonic turbulence). The full list of turbulence related simulations can be found in Table \ref{table:turbulence_table}. A projection through the core of the high mass halo for the different runs is shown in Figure \ref{fig:hm_turbulence_comparison_3pc}.\par

It is clear that even a small amount of turbulence has a dramatic effect on the evolution of the halo. In the runs with no turbulence, the halo simply collapses radially and does not fragment. When even a small amount of turbulence is added, the core becomes asymmetric and forms considerably more substructure.\par

\begin{figure*}
   \includegraphics[width=1.0\textwidth]{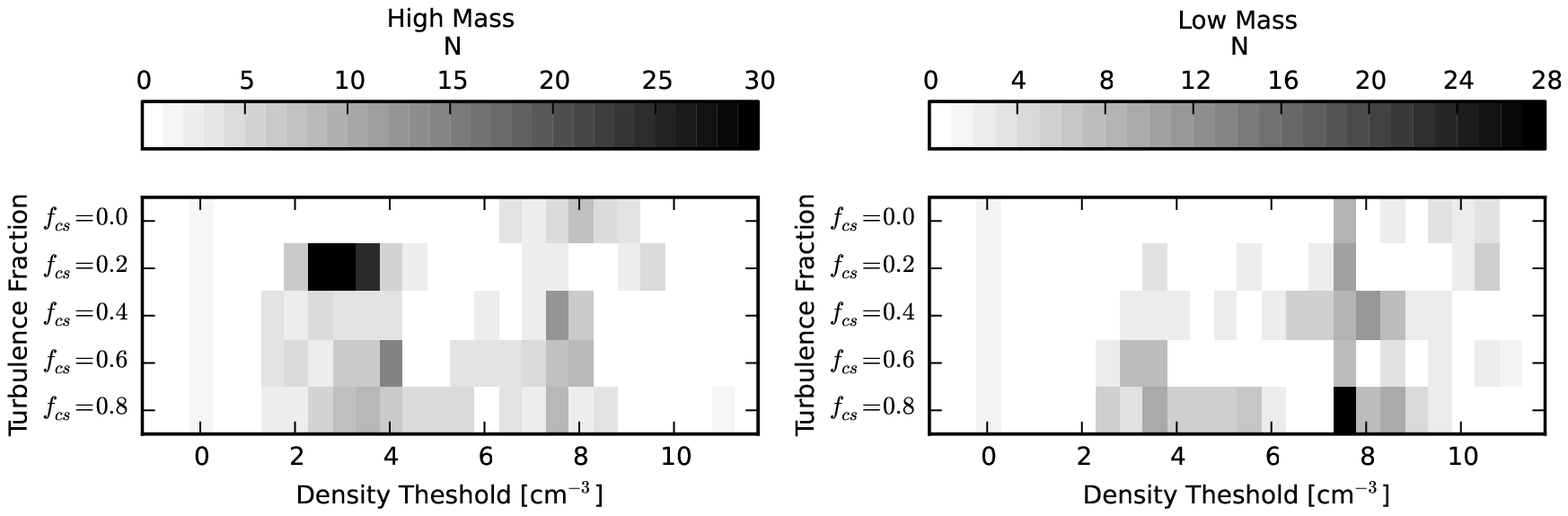}
   \centering
   \caption{The number of bound or partially bound clumps is shown for runs with with different levels of turbulence. The RMS velocity of the initial turbulent field is normalized to the fraction of the sound speed that is plotted on the y-axis. The number of clumps in each half-dex contour interval is shown above. Clump finding is performed when each run reaches a central density of $n_H=10^{10}$ \cc.}
   \label{fig:turbulence_clumps}
\end{figure*}

Once again, we attempt to quantify fragmentation by looking for potentially bound collapsing clumps within our simulation. In Figure \ref{fig:turbulence_clumps}, we show the distribution of clumps as a function of density for different levels of turbulence. While the level of fragmentation differs with the initial level of turbulence, there is not a clear relationship between the two. In fact, the most fragmentation seems to occur for intermediate levels of turbulence. We speculate that higher levels of turbulence result in substantial shock heating of the gas as the turbulence decays away, which suppresses the formation of gravitationally bound clumps and also delays the collapse of gas in the halo. This is supported by the collapse times shown in Table \ref{table:turbulence_table}, when both low and high mass halos with the highest level of turbulence take longer to collapse than their intermediate-turbulence counterparts.\par

\subsection{Dust} \label{section:varying_dust}
\begin{figure*}
   \includegraphics[width=0.95\textwidth]{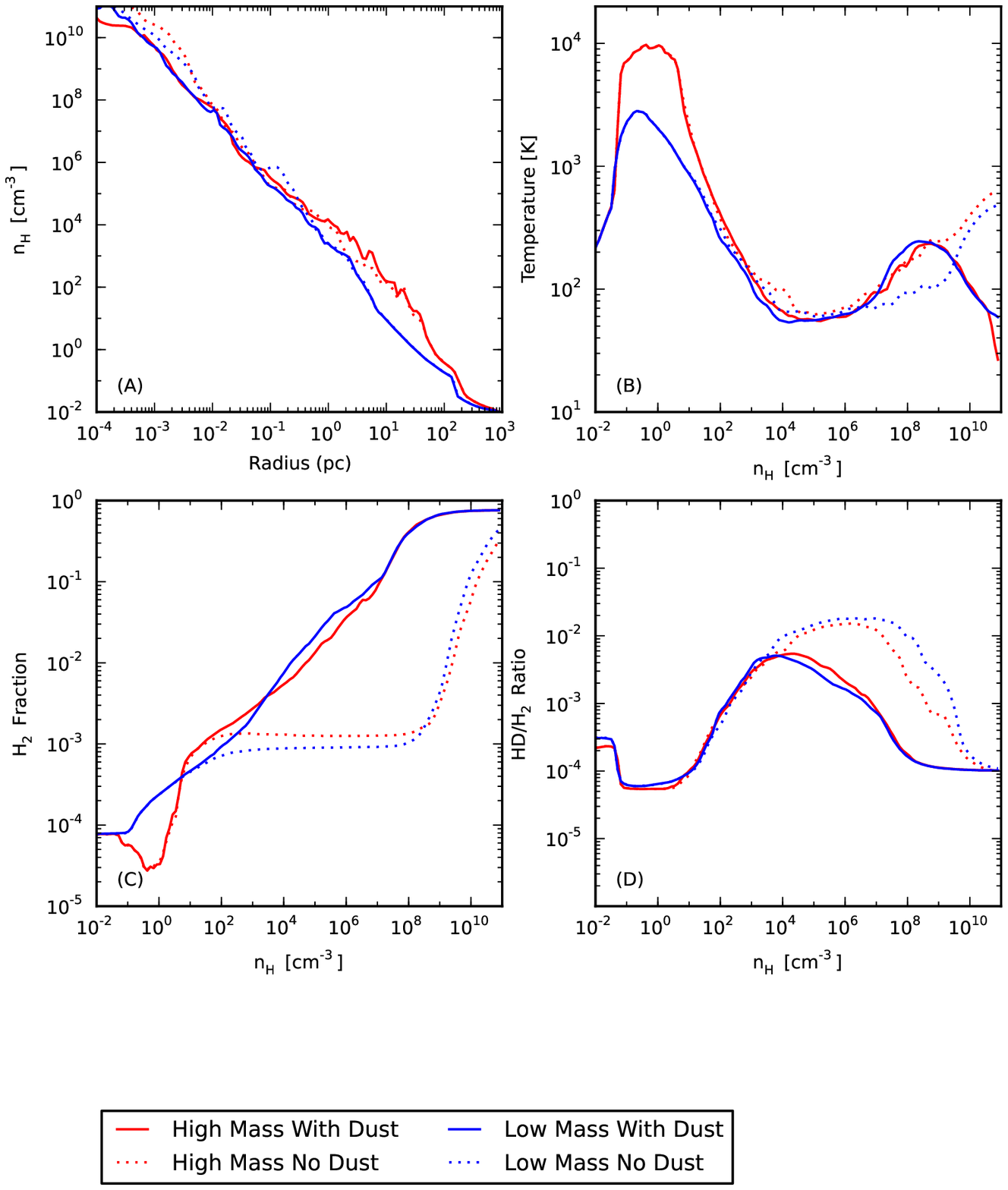}
   \centering
   \caption{The effects of dust on the physical and thermal evolution of the halo are shown for the high and low mass halo. Panel A shows the physical evolution, which is not greatly affected, as dust cooling is only important at high densities, as shown in Panel B. The primary effect of dust is to serve as a catalyst for the formation of H$_2$ and HD at low densities. The $H_2$ mass fraction is shown in Panel C. Without dust, molecules are only formed once three body reactions become important at $n_H \sim 10^8$ \cc. The addition of dust allows the gas to cool at lower densities through molecular transitions, and at high densities where the dust itself is able to radiate energy from the gas. In Panel D, the HD/H$_2$ ratio is depressed by dust as more H$_2$ is formed at low densities, decreasing the denominator.}
   \label{fig:dust_comparison}
\end{figure*}

To better understand the effects of dust on halo evolution, we ran our fiducial model without dust chemistry -- that is, assuming that all metals are in the gaseous phase. Comparisons of the physical and thermal evolution of the runs with and without dust are shown in Figure \ref{fig:dust_comparison}. Without dust, H$_2$  is not able to form in significant quantities until the onset of 3-body reactions, which occurs around a density of $n_H \sim 10^8$ \cc. This inhibits the ability of the gas to cool at low densities. Additionally, the gas does not undergo a second period of dust-driven cooling at high densities.\par

\section{Discussion} \label{section:discussion}

With the results of our model in hand, we return to the questions we set out to investigate: which physical properties of the star-forming halo affect fragmentation in low metallicity gas, and is there a true `critical metallicity' that governs the transition from a high-mass Population III stellar IMF to one that is more like the galactic IMF? We find that metallicity does have an effect on gas fragmentation at densities above $n_H \sim 10^6$ \cc, corresponding to  physical scales smaller than 0.1 pc. Our results lend tentative support to the idea of a critical metallicity found by \citet{bromm_2001} and \citet{smith_2009}, among others. However, on density scales below  $n_H \sim 10^6$ \cc, corresponding to a physical scale of greater than 0.1 pc, we do not find metallicity to have a strong impact on fragmentation. We also find that the effect of varying metallicity on the thermodynamic properties of the gas, as shown in Panel B of Figures \ref{fig:metallicity_comparison_profiles_high_mass} and \ref{fig:metallicity_comparison_profiles_low_mass}, is not as dramatic as is seen in other studies such as \citet{omukai_2005} and \citet{smith_2009}. In particular, the gas in all of our simulations is able to cool below the 200 K floor set by molecular hydrogen cooling. We explain the differences between our results and previous work as reflecting our choice of initial conditions and the physics, in particular the inclusion of dust and deuterium chemistry, in our model.\par

\subsection{The Role of HD Cooling}\label{section:choice_of_chemistry_model}
\begin{figure}
   \includegraphics[width=0.5\textwidth]{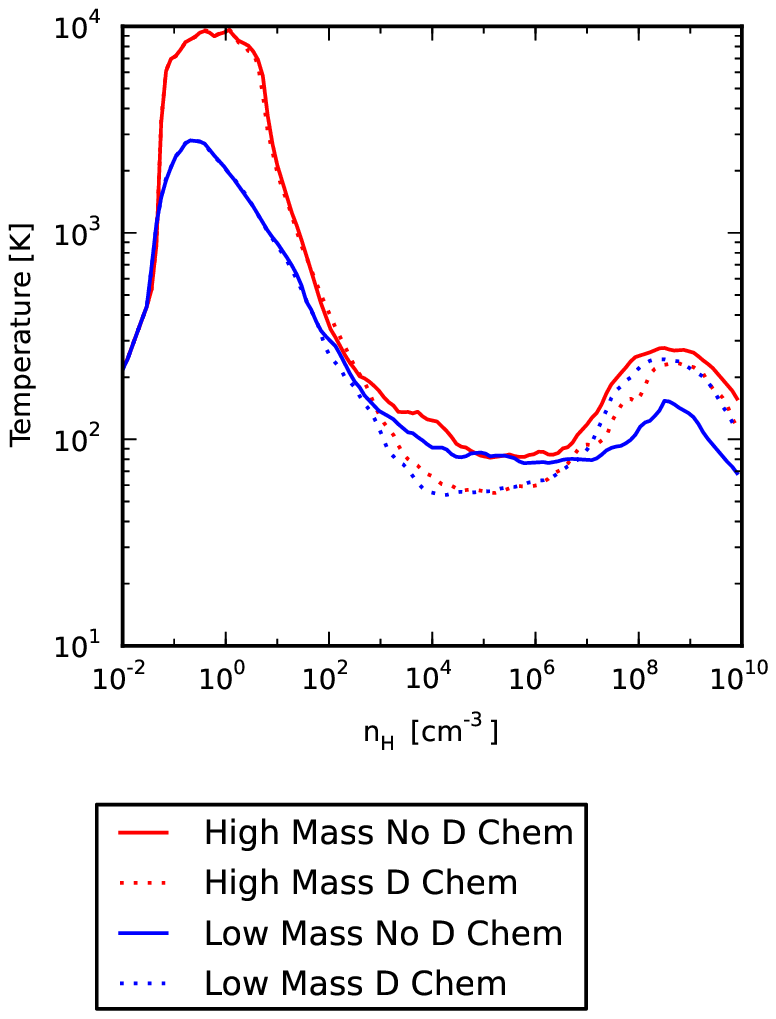}
   \centering
   \caption{The mass-weighted average temperature is shown as a function of density for simulations that include and neglect deuterium chemistry. The conditions are the same as in our high mass and low mass fiducial runs.}
   \label{fig:chemistry_temperature_density}
\end{figure}

As discussed in Section \ref{section:evolution_of_fiducial_model}, HD is a powerful coolant that can lower the temperature of the gas substantially below the limit set by H$_2$ cooling. Because the formation of HD from H$_2$ is energetically favored (see Equations \ref{equation:H2_HD_reaction} and \ref{equation:HD_H2_reaction} and the discussion that follows), enough HD can form to have a significant impact on the thermodynamic evolution of the gas. In order to confirm that HD is responsible for cooling the gas below 200 K in the absence of metals, we have rerun our fiducial model with a reduced chemical model that does not include deuterium chemistry. The results, shown in Figure \ref{fig:chemistry_temperature_density}, confirm that the inclusion of deuterium chemistry is able to lower the temperature of the gas well below the temperatures reached by H$_2$ cooling alone, even in the primordial runs. In the case of the $10^{-3}~\Zsun$ run, the gas is able to cool all the way to the CMB floor. As the combined H$_2$ and HD cooling rates dominate the metal cooling rate for metallicities below $10^{-3}~\Zsun$, there is little variation in the cooling properties of the gas at low densities, leading to little change in fragmentation at those densities as metallicity is varied (see Figures \ref{fig:metallicity_comparison_profiles_high_mass} and \ref{fig:metallicity_clumps}). For higher metallicities, the metal cooling rate dominates, leading to increasing fragmentation with increasing metallicity.\par

The importance of HD cooling in our simulations is surprising, given that other works have found HD cooling to be negligible \citep{omukai_2005, bromm_2002} in gas with low initial ionization. After investigating the factors affecting the HD cooling rate, we conclude that the amount of HD that forms is affected by the initial conditions of our simulation. Starting from a low central density, the halo is able to fully build up an accretion shock before collapsing. The collapse reaches higher temperatures, possibly ionizing the gas, and takes longer. In contrast, the one zone models used by \citet{omukai_2005} do not develop the accretion shock and assume a collapse which occurs on a dynamical timescale. \citet{bromm_2002} starts from a higher initial density and uses a top hat density profile, which again does not develop the accretion shock and allows for a faster collapse.\par

\subsection{The Role of Dust Cooling}
Recent studies \citep[e.g.,]{schneider_2006, clark_2008, dopke_2011} have found that radiation from dust grains can dominate the cooling of the gas at high ($n_H > 10^{12}$ \cc) densities for gas with metallicity above $10^{-6} \Zsun$. These simulations find support for a second metallicity threshold around $10^{-5} \Zsun$, above which dust cooling leads to a sudden drop in temperature, spurring additional fragmentation. The fragments which are formed predict a stellar IMF peaking around $1~\Msun$, consistent with modern-day star formation. As our simulations do not follow the evolution of the gas to densities above $n_H = 10^{10}$ \cc, we are not able to observe this fragmentation. However, in the simulations with metallicities high enough for dust cooling to occur at densities below $n_H\sim 10^{10}~$ \cc, we observe that this dust cooling phase can lead to significant fragmentation.  Therefore, we expect that we would have observed fragmentation in our lower metallicities runs had we carried them up to higher densities.\par

\subsection{Initial Density Profile} \label{section:varying_density_profile}

\begin{figure*}
   \includegraphics[width=1.0\textwidth]{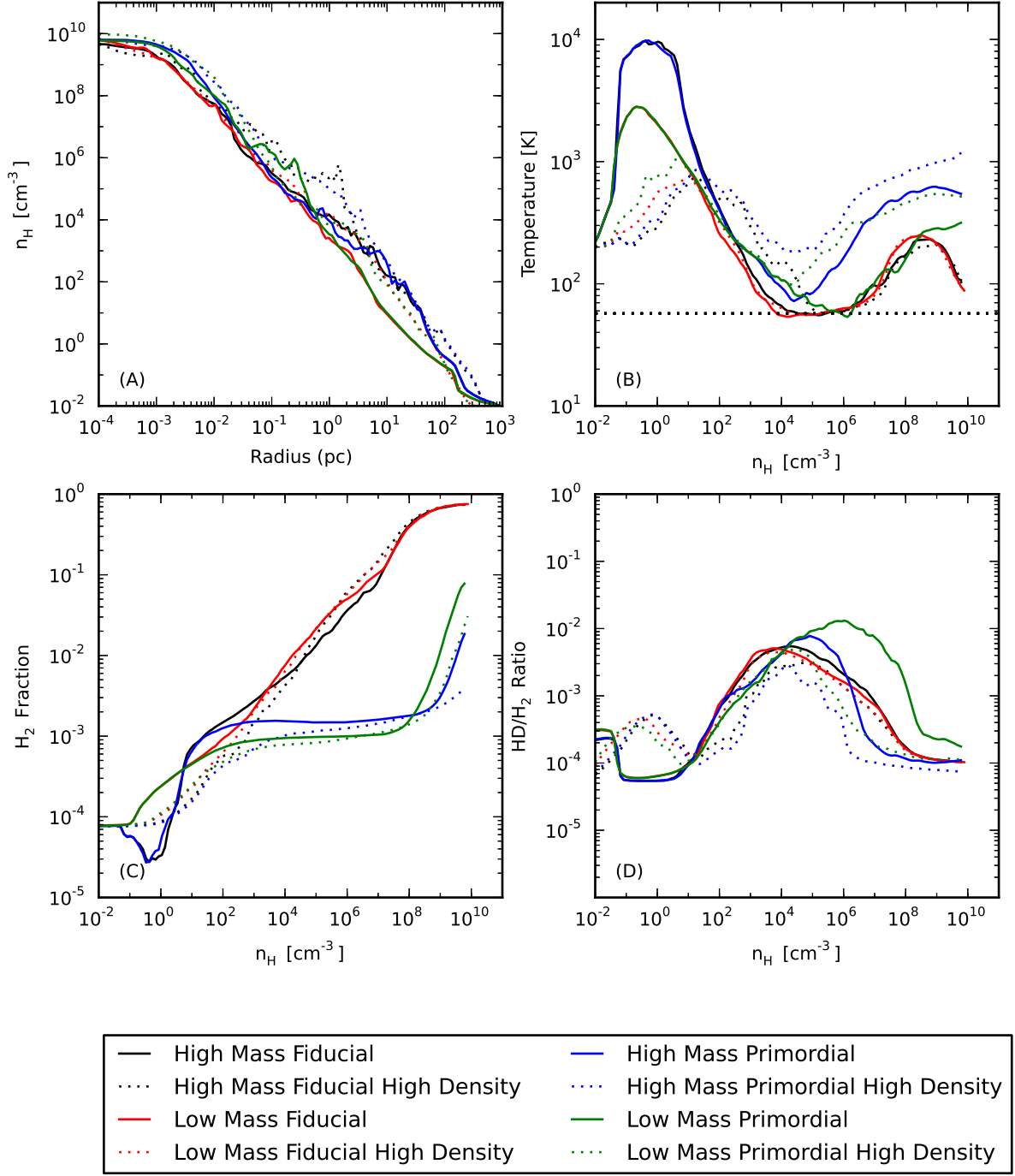}
   \centering
   \caption{The effects on the physical, thermal, and chemical evolution of the gas when the simulation is started from a higher initial gas density. Conditions are the same as in our fiducial run, except that the central gas density is $\rho_c=100$ \cc instead of $\rho_c=1$ in our fiducial model. When the central density is higher, the gas collapses before it has time to erase the imprint of the initial conditions. The accretion shock is not fully formed, resulting in no ionization and the formation of less HD at low densities. The lower HD fraction prevents the gas from cooling to the level seen in the fiducial model.}
   \label{fig:higher_density}
\end{figure*}

When we start our model from a relatively high initial density ($n_H = 10^2$ \cc), the halo begins to collapse and fragment before the gas has been fully heated by the accretion shock. This speeds up the collapse and does not give the gas enough time to build up a significant amount of HD. This in turn inhibits the gas from cooling at low densities. This is shown in Figure \ref{fig:higher_density}, which compares the physical, thermal, and chemical properties of our high and low mass fiducial models, as well as our high and low mass primordial models, with runs started from a higher central density. Divergence in the evolution of the gas is most evident in Panel b at densities below $n_H \sim 10^2$ \cc. The gas collapses before the accretion shock has fully developed, and the temperature reaches a maximum value below 1,000 K, compared to over 10,000 K for the fiducial high mass halo.\par

\subsection{Limitations of This Work}\label{section:limitations}
While our simulations attempt to accurately model the collapse and evolution of star forming halos of primordial composition and at low metallicities, our model is necessarily limited by our idealized initial setup and from the choice of physics included in our simulation. By modeling the collapse as spherically symmetric, we ignore the effects of gas accreting along filaments, which might modify the accretion shock we observe. Similarly, we model an isolated halo, which eliminates halo growth and heating due to mergers. This has been shown to affect the thermodynamic behavior of the gas \citep[e.g.,][]{oshea_2007}, but in somewhat unpredictable ways.\par

Our model makes several assumptions about the chemistry of the gas that may effect the evolution of the halo. While our assumption of uniform metallicty is likely not correct, violent relaxation in post-merger halos could efficiently mix the gas, making our approximation of uniformity appropriate. Throughout this work, we have assumed that the distribution of metals follows a scaled solar abundance. While it is quite possible that the heavy elements produced by the first generation of stars might not have had a solar abundance pattern (as has been implied by observations of metal poor stars -- see \citet{beers_2005}), what matters in our simulation is the overall cooling rate of the gas, not the details of the composition. If the composition of the gas were varied compared to solar, it would potentially change the metallicity where changes to the fragmentation become evident, but not the qualitative behavior shown in this work.\par

Throughout this work, we have assumed that the gas in the halo of interest is mostly neutral and has not been ionized by previous star formation. It is important to note that many low metallicity stars may form in halos that have hosted previous generations of star formation, meaning that our assumption of no previous ionization may not be valid in all cases. As discussed in \citet{smith_2009} and \citet{glover_2008} among others, previous ionization and subsequent recombination could affect the molecular fraction, as free electrons serve as catalysts during the molecule formation process.\par

Another source of uncertainty comes in the assumed properties of dust in our dust model. Through out this work, we have assumed dust grains with a size distribution and composition similar to grains in the solar neighborhood. If the properties of dust grains in early universe or low metallicity environments were drastically different from those in our model, our assumptions of the H$_2$ formation rate on dust grains and the rate of cooling from dust would not necessarily be valid. Until dust properties can be further constrained, the effects of dust on the formation of the first stars can not be fully understood, and our assumption of solar neighborhood like dust properties is reasonable (see \citet{omukai_2005, schneider_2006, schneider_2010} for more discussion).\par

As discussed in the preceding section, the initial conditions of the simulation are important for determining the outcome of the fragmentation process. Perhaps the largest uncertainty in this study comes in the choice of initial conditions for our low metallicity halos. Although we have attempted to base our simulations on the results of cosmological simulations, the properties of `typical' early universe star forming halos are still in the process of being constrained \citep[e.g., see][]{crosby_2013}. Future generations of semi-analytic and cosmological hydrodynamic simulations will be able to better constrain the conditions under which the Population III to metal enriched star formation transition occurred.\par

\section{Summary and Conclusions}\label{section:conclusion}
In this work, we have explored the parameter space of fragmentation in low metallicity star forming halos with the goal of better understanding the transition between metal-free and metal-enriched star formation. We have done so using the adaptive mesh hydrodynamics code Enzo, with initial conditions modeled on the results of simulations that start from cosmological initial conditions. Our simulations utilize a chemical model that includes deuterium chemistry. We have also included a dust model that tracks the formation of H$_2$ on dust grains as well as heating and cooling by dust grains. In our study, we have systematically varied the metallicity, the initial spin rate, and the level of turbulence of halos with initial dark matter masses of $10^6~\Msun$ and $10^7~\Msun$, with the aim of determining the effects of each parameter on gas evolution and fragmentation. Additionally, we have conducted simulations where we vary the physics that are included in our model and the form of our initial conditions in order to investigate how these properties affect our results.\par

We have carried out a number of simulations where $N_J$, the number of cells required to cover the local Jeans length, is varied. We find that a change in the qualitative properties of the fragmentation occurs after between $N_J=32$ and $N_J=64$. We use $N_J=64$ in our studies, but caution that increasing $N_J$ further might have non-negligible effects on the fragmentation.\par 

We conclude that varying the metallicity of the cloud has the largest impact on fragmentation, although its influence is less important than in previous works. As metallicity is increased, the gas is able to cool and collapse faster, which increases fragmentation. Above a metallicity of $10^{-4}~\Zsun$, the gas is able to fragment at higher densities, leading to the formation of substructure on sub-parsec scales and a multitude of possible star formation sites. We would likely see dust induced fragmentation at lower metallicities if we carried our simulations up to higher densities. We find tentative support for the idea of a critical metallicity, but do not see as much of a variation in evolution as has been reported in previous works. We find that the initial spin has negligible effect on fragmentation. The level of turbulence in the initial velocity field has been shown to alter the fragmentation of the cloud, but does not do so in a systematic way, with intermediate levels of turbulence typically resulting in more fragmentation than either high or low levels.\par

Our final results were found to be influenced by the initial conditions of our simulation as well as the physics included in our code, and are in generally good agreement with previous works. We found that the inclusion of deuterium chemistry alters the thermal evolution of the gas at all metallicities by allowing the gas to cool below the lower limit of H$_2$ at densities lower than the regime in which metal cooling dominates. The amount of HD that is formed and the densities where it forms is heavily dependent on the initial density profile and the subsequent evolution of the cloud. In this study we have purposely started from a low initial density so that the simulation will have time to evolve, and thus erase the details of the initial conditions. When we have started the simulation from a higher central density, the halo collapses before the gas has had time to fully form an accretion shock. The resulting collapse does not form a significant amount of HD, resulting in higher temperatures during the collapse.\par

The initial mass function of the first stars and the nature of the transition from metal free to low metallicity star formation remain open questions. As current observations cannot directly detect the first generation of stars, simulation has emerged as the main method for studying the evolution of baryons in the early universe. Semi-idealized simulations are a powerful tool for exploring the formation and evolution of the first stars, but their results can only be considered valid if the simulations include the relevant physics and initial conditions, which must be inferred from simulations based on cosmological initial conditions. Further, these calculations must be resolved numerically; inadequate spatial resolution suppresses fragmentation, thus fundamentally affecting results. These simulations in turn can benefit from semi-idealized models in order to determine what regions are most likely to host the sites of low metallicity star formation. It is our hope that with future increases in computing power and a better understanding of the conditions in the early universe, the transition from Population III to metal-enriched star formation and the history of the first stars in the universe can be fully understood.\par

\acknowledgments
\section{Acknowledgments}
This work used the Extreme Science and Engineering Discovery Environment (XSEDE), which is supported by National Science Foundation grant number OCI-1053575. All simulations were funded by XSEDE award TG-AST090040 and performed on the TACC Ranger and Stampede resources. This work was funded by the NASA ATFP program (NNX09AD80G and NNX12AC98G), the NSF AST program (AST-0908819), the LANL Institute for Geophysics and Planetary Physics, and MSU's Institute for Cyber-Enabled Research. The authors would like to thank Mike Norman, Matthew Turk, and Jeff Oishi for useful discussions. We would also like to thank Mark Voit of MSU for his support of this project. \texttt{Enzo} and \texttt{yt} are developed by a large number of independent research from numerous institutions around the world.  Their committment to open science has helped make this work possible.

\bibliographystyle{apj}
\bibliography{apj-jour,ms}

\clearpage
\begin{table}
   \centering
   \caption{Varying Metallicity}
   \begin{tabular}{c c c c}
      \hline\hline
      Run & Halo Mass (M$_\sun$) & Metallicity (Z$_\sun$) & Collapse Time \\
      \hline
      lmzmi & $10^6$ & 0.0 & 95.156 \\
      lmzm6 & $10^6$ & $10^{-6}$ & 95.070 \\
      lmzm5 & $10^6$ & $10^{-5}$ & 94.677 \\
      lmzm4 & $10^6$ & $10^{-4}$ & 93.430 \\
      lmzm3 & $10^6$ & $10^{-3}$ & 91.196 \\
      lmzm2 & $10^6$ & $10^{-2}$ & 85.805 \\
      \\

      hmzmi & $10^7$ & 0.0 & 57.229 \\
      hmzm6 & $10^7$ & $10^{-6}$ & 57.506 \\
      hmzm5 & $10^7$ & $10^{-5}$ & 56.532 \\
      hmzm4 & $10^7$ & $10^{-4}$ & 56.472 \\
      hmzm3 & $10^7$ & $10^{-3}$ & 55.360 \\
      hmzm2 & $10^7$ & $10^{-2}$ & 51.314 \\
   \end{tabular}
   \tablecomments{These are the runs performed in this work to test the effects of varying metallicity. Aside from the metallicity, all runs have the same parameters as the fiducial models. The last column gives the time in millions of years for the simulation to reach a maximum density of $n_H=10^{10}$ cm$^{-3}$.}
   \label{table:metallicity_table}
\end{table}

\begin{table}
   \centering
   \caption{Varying Jeans Refinement}
   \begin{tabular}{c c c c}
      \hline\hline
      Run & Halo Mass (M$_\sun$) & Jeans Cells & Collapse Time (Myr) \\
      \hline
      lmj4 & $10^6$ & 4 & 91.114 \\
      lmj8 & $10^6$ & 8 & 91.150 \\
      lmj16 & $10^6$ & 16 & 91.912 \\
      lmj32 & $10^6$ & 32 & 91.531 \\
      lmj64 & $10^6$ & 64 & 91.606 \\
      \\

      hmj4 & $10^7$ & 4 & 54.381 \\
      hmj8 & $10^7$ & 8 & 54.841 \\
      hmj16 & $10^7$ & 16 & 52.673 \\
      hmj32 & $10^7$ & 32 & 54.407 \\
      hmj64 & $10^7$ & 64 & 54.257 \\
   \end{tabular}
   \tablecomments{These are the runs performed in this work to test the effects of varying the Jeans refinement criteria. Other than the number of cells required to cover the Jeans length, all runs have the same parameters as the fiducial models. The last column gives the time in millions of years for the simulation to reach a maximum density of $n_H=10^{10}$ cm$^{-3}$.}
   \label{table:jeans_table}
\end{table}
   
\begin{table}
   \centering
   \caption{Varying Spin}
   \begin{tabular}{c c c c}
      \hline\hline
      Run & Halo Mass (M$_\sun$) & Spin Parameter $\lambda$ & Collapse Time (Myr) \\
      \hline
      lmsp00 & $10^6$ & 0.00 & 91.614 \\
      lmsp01 & $10^6$ & 0.01 & 91.572 \\
      lmsp03 & $10^6$ & 0.03 & 91.519 \\
      lmsp05 & $10^6$ & 0.05 & 91.603 \\
      lmsp07 & $10^6$ & 0.07 & 91.582 \\
      lmsp09 & $10^6$ & 0.09 & 91.533 \\
      \\

      hmsp00 & $10^7$ & 0.00 & 53.731 \\
      hmsp01 & $10^7$ & 0.01 & 54.054 \\
      hmsp03 & $10^7$ & 0.03 & 54.427 \\
      hmsp05 & $10^7$ & 0.05 & 54.051 \\
      hmsp07 & $10^7$ & 0.07 & 53.023 \\
      hmsp09 & $10^7$ & 0.09 & 54.302 \\
   \end{tabular}
   \tablecomments{These are the runs performed in this work to test the effects of varying the rotation, as characterized by the dimensionless spin parameter $\lambda$. Other than the spin parameter, all runs have the same parameters as the fiducial models. The last column gives the time in millions of years for the simulation to reach a maximum density of $n_H=10^{10}$ cm$^{-3}$.}
   \label{table:spin_table}
\end{table}

\begin{table}
   \centering
   \caption{Varying Turbulence}
   \begin{tabular}{c c c c}
      \hline\hline
      Run & Halo Mass (M$_\sun$) & Turbulence Factor $\lambda$ & Collapse Time (Myr) \\
      \hline
      lmt00 & $10^6$ & 0.0 & 70.278 \\
      lmt02 & $10^6$ & 0.2 & 86.686 \\
      lmt04 & $10^6$ & 0.4 & 91.599 \\
      lmt06 & $10^6$ & 0.6 & 98.232 \\
      lmt08 & $10^6$ & 0.8 & 108.922 \\
      \\

      hmt00 & $10^7$ & 0.0 & 19.760 \\
      hmt02 & $10^7$ & 0.2 & 55.234 \\
      hmt04 & $10^7$ & 0.4 & 54.087 \\
      hmt06 & $10^7$ & 0.6 & 61.464 \\
      hmt08 & $10^7$ & 0.8 & 62.704 \\
   \end{tabular}
   \tablecomments{These are the runs performed in this work to test the effects of varying the degree of turbulence. The RMS of the initial turbulent velocity field is normalized to some fraction of the halo sound speed. This is shown in the third column. Other than the degree of turbulence, all runs have the same parameters as the fiducial models. The last column gives the time in millions of years for the simulation to reach a maximum density of $n_H=10^{10}$ cm$^{-3}$.}
   \label{table:turbulence_table}
\end{table}

\begin{table}
   \centering
   \caption{Other Runs}
   \begin{tabular}{c c c c}
      \hline\hline
      Run & Halo Mass (M$_\sun$) & Collapse Time (Myr) & Description\\
      \hline
      lmnd & $10^6$ & 92.711 & No Dust \\
      hmnd & $10^7$ & 55.270 & No Dust \\
      \\

      lmm2 & $10^6$ & 93.013 & Reduced Chemical Network \\
      hmm2 & $10^7$ & 55.571 & Reduced Chemical Network \\
      \\

      hmhd  & $10^7$ & 4.651 & Higher Initial Density \\
      hmhdp & $10^7$ & 5.677 & Higher Initial Density (Primordial) \\
      lmhd  & $10^6$ & 12.315 & Higher Initial Density \\
      lmhdp & $10^6$ & 16.300 & Higher Initial Density (Primordial) \\
   \end{tabular}
   \tablecomments{This table shows additional runs performed in this paper. In all cases, the parameters are the same as those of our fiducial model unless otherwise noted. In the first set of runs, we do not include dust chemistry. In the second set of runs, we use a reduced chemical model which does not include Deuterium chemistry. In the third set of runs, we start with an initial baryon density 100 times higher than in our fiducial model.}
   \label{table:misc_runs_table}
\end{table}

\end{document}